%% file: mqcle_hermite_closure.tex
\documentclass[aip,jcp,amsmath,amssymb,reprint,floatfix,nofootinbib]{revtex4-2}

\usepackage{graphicx}
\usepackage{bm}
\usepackage{color}
\usepackage{tikz}
\usepackage{url}
\usetikzlibrary{arrows.meta,positioning,calc}

\newcommand{\bh}{\bar h}
\newcommand{\cL}{\mathcal L}
\newcommand{\cX}{\mathcal X}
\newcommand{\cV}{\mathcal V}

\newcommand{\cT}{\mathcal T}
\newcommand{\HH}{\mathbb H}
\newcommand{\JJ}{\mathbb J}
\newcommand{\PB}{\mathrm{PB}}
\newcommand{\He}{\mathrm{He}}
\newcommand{\rt}{\widetilde\rho}
\newcommand{\n}{\bm n}
\newcommand{\m}{\bm m}
\newcommand{\e}{\bm e}
\newcommand{\cc}{\bm c}
\newcommand{\Tr}{\operatorname{Tr}}
\newcommand{\one}{\mathchoice{\openone}{\openone}{\scalebox{0.72}{$\openone$}}{\scalebox{0.55}{$\openone$}}}
\newcommand{\wick}[1]{\mathopen{:}#1\mathclose{:}}

\newsavebox{\ladderbox}

\input{paper_values}

\begin{document}

\title{Exact Gaussian--Hermite closure of the mapping quantum--classical
Liouville equation}

\author{Sahand Nikzat}
\email{sahand.nikzat@mail.utoronto.ca}
\thanks{ORCID: \url{https://orcid.org/0009-0000-4224-4482}}
\affiliation{Chemical Physics Theory Group, Department of Chemistry,
University of Toronto, 80 St.\ George Street, Toronto, Ontario M5S 3H6,
Canada}
\author{Ali Nassimi}
\email{anasimi@gmail.com}
\thanks{ORCID: \url{https://orcid.org/0000-0002-2639-0962}}
\affiliation{Independent Researcher, St. John's, Newfoundland and Labrador,
Canada}
\author{Rony Modak Parai}
\email{ronyparai176@gmail.com}
\thanks{ORCID: \url{https://orcid.org/0009-0000-9785-6216}}
\affiliation{Chemical Physics Theory Group, Department of Chemistry,
University of Toronto, 80 St.\ George Street, Toronto, Ontario M5S 3H6,
Canada}

\date{October 11, 2026}

\begin{abstract}
In the Meyer–Miller–Stock–Thoss mapping representation, the Liouville operator of the quantum–classical Liouville equation contains an excess coupling term in addition to the Poisson bracket. Neglecting this term yields the Poisson bracket mapping equation (PBME). We expand the mapping dependence of the density in Hermite functions about the Gaussian factor common to all physical mapping densities. In this basis, the excess coupling term acts purely as a raising operator and exactly cancels the raising contribution generated by the back-reaction force. Consequently, the exact dynamics does not generate higher Hermite orders and closes on a finite, sparse system of equations for the coefficient fields. Within the physical sector, these equations reduce to the subsystem quantum–classical Liouville equation. In contrast, the PBME retains the raising contribution of the force, modifying the back reaction within the physical sector while continuously generating higher Hermite orders. For Tully’s dual avoided crossing model, direct propagation of the coefficient equations confirms this finite closure and agrees with independent solutions of the subsystem equation. Comparison with converged full quantum dynamics further shows that, across all regimes examined, the PBME is less accurate than the quantum–classical description for populations, coherences, and momentum distributions.
\end{abstract}

\maketitle

\section{Introduction}

In nonadiabatic processes in condensed phases and large molecules, a small
electronic subsystem is coupled to an environment too large to treat
quantum mechanically.\cite{kapral2006,kapral2015}  The quantum--classical
Liouville equation (QCLE) retains a quantum description of the subsystem
while representing the environment by phase-space variables obtained through
a partial Wigner transform.  The QCLE is exact for a quadratic scalar bath
Hamiltonian with subsystem--bath coupling linear in the bath variables;
this includes a subsystem bilinearly coupled to a harmonic
bath.\cite{kim2008,kapral2015}  In general, the
equation is difficult to solve: the back-reaction term couples the subsystem
density to derivatives with respect to bath momentum, and common
adiabatic-basis trajectory formulations represent the resulting dynamics
using surface hops accompanied by momentum jumps.\cite{kapral1999,mackernan2008}
The mapping formalism of Meyer, Miller, Stock, and Thoss represents each
discrete state by a harmonic oscillator and identifies the physical subsystem
with the subspace in which exactly one oscillator is excited.  In this
representation, both subsystem and bath are described using continuous
phase-space variables.\cite{meyer1979,stock1997,thoss1999,muller1998}  Kim,
Nassimi, and Kapral expressed the QCLE in mapping variables.  The resulting
Liouville operator has two contributions: a Poisson bracket on the combined
mapping and bath phase space, and an excess-coupling term containing second
derivatives with respect to mapping variables and a first derivative with
respect to bath momentum.\cite{kim2008}  Retaining only the Poisson-bracket
term gives the Poisson bracket mapping equation (PBME), which can be
propagated by independent Newtonian trajectories.\cite{kelly2012}
Related trajectory approaches include linearized and partially linearized
mapping methods,\cite{bonella2005,huo2011,gao2020} symmetrical windowing of
quasiclassical mapping trajectories,\cite{cotton2013} and spin
mapping.\cite{runeson2019}
The excess-coupling term was shown by Nassimi, Bonella, and Kapral to carry
part of the subsystem's back reaction on the bath.\cite{nassimi2010}  Kelly
\emph{et al.} proved that the full mapping QCLE operator commutes with the
projector onto the physical mapping states, so exact dynamics initialized in
that subspace remains there; the Poisson-bracket operator does not preserve
this subspace.\cite{kelly2012}  Unphysical PBME features, including negative
populations and trajectories evolving on inverted potential surfaces, are
well documented and have motivated several modifications.
\cite{kelly2012,kimrhee2014,kimleerhee2014,kimrhee2020}
Other approaches treat the mapping dynamics differently, through
approximate forward--backward trajectory solutions of the mapping
QCLE\cite{hsieh2012,hsieh2013,braver2020} or semiclassical approximations to
the exact mapping propagator.\cite{hele2016,church2018}
We study the excess-coupling term as an operator on the full mapping density,
including densities outside the physical sector.  Every physical
mapping density is a common Gaussian multiplied
by a polynomial of degree at most two in the mapping variables,\cite{kelly2012}
so a Hermite expansion about that Gaussian provides a natural description.
Hermite moment hierarchies of this kind are familiar from kinetic
theory,\cite{grad1949,li2021} where they generally do not close and finite
truncations are approximations.
We show that the Hermite hierarchy for the mapping QCLE closes exactly
because of the excess-coupling term.  In a Gaussian-weighted Hermite basis,
this term raises each Hermite component by two orders.  The mapping-dependent
part of the back-reaction force contains the same raising contribution with
the opposite sign.  The mapping flow preserves Hermite order, and bath
advection does not act on the mapping variables.  The two raising
contributions therefore cancel for every Hermite function, not only for
physical densities.  Consequently, every finite Hermite space is invariant
under the full mapping dynamics.  In the physical sector, this cancellation
recovers results of Nassimi \emph{et al.} and Kelly \emph{et al.};
\cite{nassimi2010,kelly2012} more generally, it yields a finite set of sparse
equations for Hermite coefficient fields of any initial density with finite
Hermite support.
The PBME differs from the full mapping QCLE by the uncancelled raising part
of the force.  Acting on a physical density, this term raises the order-zero
component into the physical sector, thereby changing the back reaction there,
and raises the order-two components to order four, outside the physical
sector.  Repeated applications can continue to populate higher Hermite
orders.
We test these results on Tully's dual avoided crossing\cite{tully1990} over a
range of incident momenta.\cite{nikzat2021}  We propagate the coefficient equations
directly in Hermite bases large enough to reveal any failure of the predicted
closure and compare them with two independent solvers of the subsystem QCLE,
one based on the Strang splitting commonly used for this purpose.  Converged
full quantum dynamics provides a reference for assessing the quantum--classical
approximation.  For the parameters studied, the comparison separates the
quantum--classical error from the additional error introduced by the PBME and
finds the latter to be larger in the populations, coherences, and momentum
distributions.  In phase space, both descriptions also exhibit density-level
discrepancies that are hidden when the densities are integrated to obtain
populations.  Finally, we test whether the fourth-order defect, which can be
computed directly from a QCLE solution, is a useful indicator of PBME error.
Section~\ref{sec:theory} develops the theory, Sec.~\ref{sec:methods}
describes the model and the calculations, Sec.~\ref{sec:results} presents the
results, and Sec.~\ref{sec:discussion} discusses their implications and
scope.
Derivations, symbolic verifications and additional results are collected in
the supplementary material.

\section{Theory}

\label{sec:theory}

\subsection{Quantum--classical Liouville equation and its mapping form}

\label{sec:mqcle}
Consider a quantum subsystem with $N$ states coupled to a bath coordinate $R$
with momentum $P$ and mass $M$.  For several bath coordinates, the equations
below acquire sums over $R_\alpha$ and $P_\alpha$.  The partial Wigner
transform over the bath,
\[
\hat\rho_W(R,P,t)=\frac{1}{2\pi\hbar}\int dz\,e^{iPz/\hbar}
\bigl\langle R-\tfrac z2\big|\hat\rho(t)\big|R+\tfrac z2\bigr\rangle,
\]
turns the density operator into an operator in the subsystem space that
depends on the bath phase-space point.  Its evolution is given by the QCLE,\cite{kapral1999,kapral2006}
\begin{equation}
\begin{split}
\partial_t\hat\rho_W={}&-i\hat\cL\,\hat\rho_W\\
={}&-\frac{i}{\hbar}\bigl[\hat H_W,\hat\rho_W\bigr]\\
&+\frac12\Bigl(\bigl\{\hat H_W,\hat\rho_W\bigr\}-\bigl\{\hat\rho_W,\hat H_W\bigr\}\Bigr),
\end{split}
\label{eq:qcle-op}
\end{equation}
which defines the quantum--classical Liouville operator $i\hat\cL$ in the
sign convention used here.  The partially Wigner-transformed Hamiltonian is
$\hat H_W=P^2/2M+V_b(R)+\hat h(R)$, where $V_b$ is the bath potential and
$\hat h(R)$ is the subsystem Hamiltonian, including its coupling to $R$.
The operator-valued Poisson bracket is
$\{\hat A,\hat B\}=\partial_R\hat A\,\partial_P\hat B-\partial_P\hat A\,\partial_R\hat B$;
operator products retain their written order.  Equation~\eqref{eq:qcle-op}
is exact for a quadratic scalar bath Hamiltonian and subsystem--bath
coupling linear in the bath variables.\cite{kim2008,kapral2015}  More generally,
it is the first-order truncation of the bath Moyal expansion.  In a diabatic
basis $\{|\lambda\rangle\}$, which does not depend on
$R$, let $h(R)$ and $\rho_W(R,P,t)$ be the $N\times N$ matrices of $\hat h$
and $\hat\rho_W$.  Equation~\eqref{eq:qcle-op} then reads
\begin{equation}
\begin{split}
\partial_t\rho_W={}&-\frac{i}{\hbar}\bigl[h,\rho_W\bigr]
-\frac{P}{M}\,\partial_R\rho_W\\
&+\frac12\bigl(\partial_RV\,\partial_P\rho_W+\partial_P\rho_W\,\partial_RV\bigr),
\end{split}
\label{eq:qcle}
\end{equation}
with $V=V_b\one+h$, the total potential matrix.  The last term accounts for
the forces acting on the bath.  The contribution from $\partial_R V_b$
comes from the bath potential, while that from $\partial_R h$ describes
the subsystem's back reaction through the coupling.\cite{nassimi2010}
In the mapping representation the state $\lambda$ becomes a singly excited
oscillator with coordinate $r_\lambda$ and momentum $p_\lambda$; the vector
$x$ collects the $N$ coordinates followed by the $N$ momenta.  The Wigner
transform over the oscillator coordinates turns $\hat H_W$ into the mapping
Hamiltonian\cite{meyer1979,stock1997,kim2008}
\begin{equation}
H_m=\frac{P^2}{2M}+V_b(R)+\frac{1}{2\hbar}\Bigl[x^{\mathsf T}\HH(R)\,x
-\hbar\,\Tr h(R)\Bigr],
\label{eq:Hm}
\end{equation}
where $\HH$ is the real symmetric $2N\times2N$ form of $h$: for $h=A+iB$ with
$A$ real symmetric and $B$ real antisymmetric,
$\HH=\bigl(\begin{smallmatrix}A&-B\\ B&A\end{smallmatrix}\bigr)$.  For a real
symmetric $h$, $x^{\mathsf T}\HH x=\sum_{\lambda\lambda'}h_{\lambda\lambda'}
(r_\lambda r_{\lambda'}+p_\lambda p_{\lambda'})$, which is the form used in
Ref.~\onlinecite{kim2008}.
A subsystem density matrix is represented by the physical mapping density\cite{kelly2012}
\begin{equation}
\begin{gathered}
\rho^{\rm phys}(x,R,P)=\phi(x)\sum_{\lambda\lambda'}\Pi_{\lambda\lambda'}(x)\,
\rho_W^{\lambda'\lambda}(R,P),\\
\Pi_{\lambda\lambda'}=\frac{2}{\hbar}\Bigl[r_\lambda r_{\lambda'}
+p_\lambda p_{\lambda'}-i\bigl(r_\lambda p_{\lambda'}-r_{\lambda'}p_\lambda\bigr)
-\frac{\hbar}{2}\delta_{\lambda\lambda'}\Bigr],
\end{gathered}
\label{eq:rhophys}
\end{equation}
with $\phi(x)=(\pi\hbar)^{-N}e^{-x\cdot x/\hbar}$; $\phi\,\Pi_{\lambda\lambda'}$
is the function $g_{\lambda\lambda'}$ of Ref.~\onlinecite{kelly2012}, and
$\rho^{\rm phys}$ is the projected mapping density of that work.  With these
normalizations the phase-space average of the mapping symbol of any subsystem
operator $\hat O$ over $\rho^{\rm phys}$ is $\Tr(\hat O\rho_W)$
(supplementary material, Sec.~S1).  We call the span of the $N^2$ functions
$\phi\,\Pi_{\lambda\lambda'}$ the physical sector.
A mapping density $\rho_m(x,R,P,t)$ evolves according to the mapping
QCLE,\cite{kim2008,nassimi2010,kelly2012}
\begin{equation}
\partial_t\rho_m=-i\cL_m\rho_m,\qquad i\cL_m=i\cL_m^{\PB}+i\cL_m'.
\label{eq:mqcle-full}
\end{equation}
The Poisson-bracket contribution acts on the joint mapping--bath phase space
$\cX=\{x,R,P\}$ as
\begin{equation}
i\cL_m^{\PB}\rho_m=-\{H_m,\rho_m\}_{\cX}.
\label{eq:mqcle-pb}
\end{equation}
The excess-coupling operator contains second derivatives with respect to the
mapping variables and a first derivative with respect to bath momentum:
\begin{equation}
i\cL_m'=\frac{\hbar}{8}\sum_{ij}\partial_R\HH_{ij}\,
\frac{\partial^2}{\partial x_i\partial x_j}\frac{\partial}{\partial P},
\label{eq:mqcle-excess}
\end{equation}
where $\{\cdot,\cdot\}_{\cX}$ is the Poisson bracket in $\cX$ and $i,j$
range over the $2N$ mapping coordinates; the derivatives act on the density to
their right.  The mapping-space Wigner transform turns the operator products in the bath
Poisson bracket into Moyal products.  The symmetrized mapping-space product
has an even-derivative expansion; because $H_m$ is quadratic in $x$, it
terminates at second order and yields the excess-coupling term in
Eq.~\eqref{eq:mqcle-excess} (supplementary material, Sec.~S2).  For real
symmetric $h$, this is the operator given in Refs.~\onlinecite{kim2008} and
\onlinecite{kelly2012}; for complex Hermitian $h$, the block-matrix form
written using $\HH$ is required.  Dynamical variables evolve with the
adjoint convention $dB_m/dt=i\cL_mB_m$.\cite{kim2008,kelly2012}
Substituting Eq.~\eqref{eq:rhophys} into the full mapping QCLE,
Eq.~\eqref{eq:mqcle-full}, reproduces
Eq.~\eqref{eq:qcle} exactly.  We have verified this identity by computer
algebra for real and complex Hermitian $h$ (supplementary material,
Sec.~S5).  Dropping $i\cL_m'$ from Eq.~\eqref{eq:mqcle-full} gives the PBME,
$\partial_t\rho_m=-i\cL_m^{\PB}\rho_m$.  Below, we call $i\cL_m$ the full
Liouville operator, exact with respect to the QCLE, and $i\cL_m^{\PB}$ its
Poisson-bracket approximation.
The scalar part of $h$ can be absorbed into the bath potential.  Write
$h=\bh+N^{-1}(\Tr h)\one$, replace $h$ by $\bh$, and replace $V_b$ by
$V_b+N^{-1}\Tr h$.  This leaves the total potential $V_b\one+h$ and hence
Eq.~\eqref{eq:qcle} unchanged.  The action of the full mapping operator
$i\cL_m$ on physical densities is also unchanged, provided the same
traceless matrix is used consistently in $H_m$ and $i\cL_m'$ (Sec.~S2).
This redistribution changes $i\cL_m^{\PB}$ and $i\cL_m'$ separately, and
therefore changes the PBME.  We use the traceless convention.  Kelly
\emph{et al.} found that this choice reduced the occurrence of inverted
potential surfaces in their model calculations.\cite{kelly2012}

\subsection{Gaussian--Hermite representation}

\label{sec:hermite}
The standardized variables $z=(2/\hbar)^{1/2}x$ turn $\phi$ into the unit
Gaussian, $\phi(x)\,dx=\varphi(z)\,dz$ with
$\varphi(z)=(2\pi)^{-N}e^{-z\cdot z/2}$.  We expand the mapping dependence of
the density $\rt(z,R,P,t)$ in these variables, defined by $\rt\,dz=\rho_m\,dx$,
as
\begin{equation}
\rt(z,R,P,t)=\varphi(z)\sum_{\n}c_{\n}(R,P,t)\,\He_{\n}(z),
\label{eq:expansion}
\end{equation}
where $\He_{\n}(z)=\prod_i\He_{n_i}(z_i)$ is a product of probabilists'
Hermite polynomials, $\n=(n_1,\dots,n_{2N})$, and $|\n|=\sum_in_i$ is its
order.  The functions $\varphi\He_{\n}$ are orthogonal,
$\int dz\,\varphi\He_{\m}\He_{\n}=\n!\,\delta_{\m\n}$ with
$\n!=\prod_in_i!$.  Only the mapping dependence is expanded: the coefficients
are fields on the bath phase space.  We denote by $\cV_K$ the densities with
$c_{\n}=0$ for $|\n|>K$.
In these variables
$\Pi_{\lambda\lambda'}=z_\lambda z_{\lambda'}+z_{N+\lambda}z_{N+\lambda'}
-i(z_\lambda z_{N+\lambda'}-z_{\lambda'}z_{N+\lambda})-\delta_{\lambda\lambda'}$
is an even polynomial of second degree, so every physical density lies in
$\cV_2$, and the lowest coefficient is the trace, $c_{\bm 0}=\Tr\rho_W$.  For
$N=2$ the even part of $\cV_2$ has $1+N(2N+1)=11$ coefficient fields, and the
physical densities span a four-dimensional subspace of it.  Invariance of $\cV_2$ is therefore a weaker
statement than invariance of the physical sector, and the two are kept apart
below.
\subsection{The excess coupling term is a raising operator}
\label{sec:raising}
\begin{figure*}[t]
\centering
\begin{minipage}[c]{0.56\textwidth}
\centering
\sbox{\ladderbox}{\begin{tikzpicture}[>={Latex[length=1.8mm]},font=\small,x=1cm,y=1cm]
  \node[font=\bfseries] at (-0.55,4.35) {(a)};
  \foreach \X/\lab in {0/{full, $i\cL_m$},4.35/{Poisson bracket, $i\cL_m^{\PB}$}}{
    \begin{scope}[xshift=\X cm]
      \foreach \k/\y in {0/0,2/1.2,4/2.4,6/3.6}{
        \draw[gray!55] (0,\y) -- (3.7,\y);
        \node[gray!60!black,anchor=east,font=\footnotesize] at (0,\y) {$\k$};}
      \node[anchor=south] at (1.85,3.85) {\lab};
      \fill[blue!10] (0.12,-0.2) rectangle (0.95,1.4);
      \node[blue!50!black,font=\scriptsize,rotate=90] at (0.3,0.6) {physical};
    \end{scope}}
  \node[rotate=90,gray!60!black,font=\footnotesize] at (-0.55,1.8) {Hermite order};
  \draw[->,thick] (0.6,1.33) arc[start angle=200,end angle=-20,radius=0.22];
  \draw[->,thick,green!45!black] (1.25,1.12) -- (1.25,0.08);
  \draw[->,thick,red!70!black] (1.85,1.28) -- (1.85,2.32);
  \draw[->,thick,red!70!black] (2.55,1.28) -- (2.55,2.32);
  \node[red!70!black,anchor=east,font=\footnotesize] at (1.82,1.8) {$+\tfrac14$};
  \node[red!70!black,anchor=west,font=\footnotesize] at (2.58,1.8) {$-\tfrac14$};
  \draw[red!70!black,thick] (1.7,2.62) -- (2.7,2.62);
  \node[anchor=south,font=\scriptsize,align=center] at (2.2,2.62) {force $+$ excess\\[-1pt] cancel};
  \begin{scope}[xshift=4.35cm]
    \draw[->,thick] (0.6,1.33) arc[start angle=200,end angle=-20,radius=0.22];
    \draw[->,thick,green!45!black] (1.25,1.12) -- (1.25,0.08);
    \draw[->,thick,red!70!black] (1.9,0.08) -- (1.9,1.12);
    \draw[->,thick,red!70!black] (1.9,1.28) -- (1.9,2.32);
    \draw[->,thick,red!70!black,dashed] (1.9,2.48) -- (1.9,3.52);
    \node[anchor=west,font=\scriptsize,align=left] at (2.0,0.6) {back reaction\\[-1pt] altered, Eq.~\eqref{eq:inframe}};
    \node[anchor=west,font=\scriptsize,align=left] at (2.0,1.8) {leak, Eq.~\eqref{eq:quartic}};
    \node[anchor=west,font=\scriptsize] at (2.0,3.0) {and so on};
  \end{scope}
\end{tikzpicture}}%
\ifdim\wd\ladderbox>\linewidth
  \resizebox{\linewidth}{!}{\usebox{\ladderbox}}%
\else
  \usebox{\ladderbox}%
\fi
\end{minipage}\hfill
\begin{minipage}[c]{0.42\textwidth}
\centering
\includegraphics[width=\linewidth]{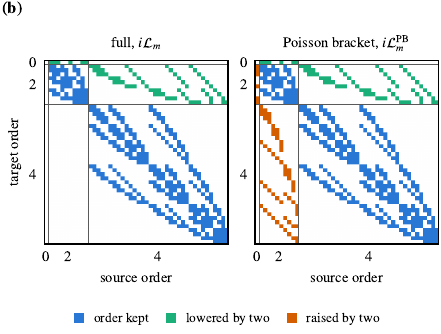}
\end{minipage}
\caption{\label{fig:ladder}Couplings between Hermite orders.  (a) Under the full
Liouville operator the raising contributions of the force ($+\tfrac14$) and of
the excess coupling term ($-\tfrac14$) cancel, so every coupling keeps the
order (loop) or lowers it by two; in the Poisson bracket approximation the
raising part survives.  (b) Nonzero blocks of the coefficient operators in
Eqs.~\eqref{eq:coeff} and \eqref{eq:PBcoeff} for $N=2$, coloured by the
change of order.}
\end{figure*}

Differentiation raises the Hermite order,
$\partial_{z_i}(\varphi\He_{\n})=-\varphi\He_{\n+\e_i}$, so that
\begin{equation}
\frac{\partial^2}{\partial z_i\partial z_j}\bigl(\varphi\He_{\n}\bigr)
=\varphi\He_{\n+\e_i+\e_j},
\label{eq:raise}
\end{equation}
while multiplication by $z_i$ both raises and lowers it,
$z_i\He_{\n}=\He_{\n+\e_i}+n_i\He_{\n-\e_i}$.  In the variables $z$ the
excess term is
$i\cL_m'=\tfrac14\sum_{ij}\partial_R\HH_{ij}\,\partial_{z_i}\partial_{z_j}\partial_P$,
and by Eq.~\eqref{eq:raise} its contribution to the time derivative of each
Hermite component is
\begin{equation}
-i\cL_m'\bigl[c_{\n}\,\varphi\He_{\n}\bigr]
=-\frac14\sum_{ij}\partial_R\HH_{ij}\,\bigl(\partial_Pc_{\n}\bigr)\,
\varphi\He_{\n+\e_i+\e_j}.
\label{eq:excess-raise}
\end{equation}
The excess coupling term therefore moves every component up by two orders
and does nothing else.
The Poisson bracket operator $i\cL_m^{\PB}$ has three parts.  The mapping part generates the flow
$\dot x=Kx$ with $K=\hbar^{-1}\JJ\HH$ and
$\JJ=\bigl(\begin{smallmatrix}0&\one\\-\one&0\end{smallmatrix}\bigr)$.
Since $\HH$ commutes with $\JJ$ for every Hermitian $h$, $K$ is
antisymmetric.  The resulting flow is a rotation of mapping space: it
preserves the Gaussian and each subspace of fixed total Hermite order, while
mixing basis functions within that subspace.  The bath advection $-(P/M)\partial_R$ does not
act on $z$.  The bath-momentum term $\partial_RH_m\,\partial_P\rho_m$ of
$-i\cL_m^{\PB}\rho_m$ multiplies the
momentum derivative of the density by the potential gradient
$\partial_RH_m=\partial_RV_b+\tfrac14\sum_{ij}\partial_R\HH_{ij}(z_iz_j-\delta_{ij})$,
and applying the recurrence for $z_i\He_{\n}$ twice gives
\begin{equation}
(z_iz_j-\delta_{ij})\He_{\n}=\He_{\n+\e_i+\e_j}+\cT_{ij}\He_{\n},
\label{eq:force-split}
\end{equation}
with
\begin{equation}
\begin{split}
\cT_{ij}\He_{\n}={}&n_j\He_{\n+\e_i-\e_j}+n_i\He_{\n-\e_i+\e_j}\\
&+n_i(n_j-\delta_{ij})\He_{\n-\e_i-\e_j}.
\end{split}
\label{eq:T}
\end{equation}
Terms in Eq.~\eqref{eq:T} with a negative Hermite index are understood to
vanish.  The first term of Eq.~\eqref{eq:force-split} is the raising part of
the force.
It enters the time derivative with the coefficient $+\tfrac14\partial_R\HH_{ij}$
and the same Hermite function as Eq.~\eqref{eq:excess-raise}, which carries
$-\tfrac14\partial_R\HH_{ij}$.  The two cancel for every $\n$.  What remains of
the force is $\cT_{ij}$, which keeps the order or lowers it by two.
The full Liouville operator therefore never raises the Hermite order.
Substituting the Hermite expansion into the full mapping QCLE and collecting
coefficients in $\cc(R,P,t)$ gives
\begin{equation}
\partial_t\cc=-\frac{P}{M}\,\partial_R\cc+\mathsf R(R)\,\cc
+\mathsf G(R)\,\partial_P\cc,
\label{eq:coeff}
\end{equation}
with
\begin{equation}
\mathsf R=-\sum_{ij}K_{ij}\,\mathsf N^{ij},\qquad
\mathsf G=\partial_RV_b\,\mathsf 1+\frac14\sum_{ij}\partial_R\HH_{ij}\,
\mathsf T^{ij}.
\label{eq:RG}
\end{equation}
The matrices $\mathsf N^{ij}_{\m\n}=n_i\,\delta_{\m,\n-\e_i+\e_j}$ and
$\mathsf T^{ij}$, whose elements are the coefficients in Eq.~\eqref{eq:T}, are
sparse integer matrices fixed by the Hermite basis; the model enters only
through the fields $K(R)$, $\partial_R\HH(R)$ and $\partial_RV_b(R)$.  We have
checked Eqs.~\eqref{eq:coeff} and \eqref{eq:RG} against the full mapping
QCLE by computer algebra on every Hermite function of order up to four
(supplementary material, Sec.~S5).  The
rotation couples $\n$ to $\n-\e_i+\e_j$ and keeps the order; the force couples
it to the same targets and to $\n-\e_i-\e_j$.  The equation for a
coefficient of order $k$ therefore depends only on
coefficients of orders $k$ and $k+2$.  Hermite-order parity is conserved, and
every $\cV_K$ is invariant.  Thus, for the full mapping QCLE, a density
initially supported through order $K$ can be propagated with an order-$K$
truncation without approximation.
Figure~\ref{fig:ladder} shows the resulting coupling pattern for $N=2$ in the
basis of all even orders up to four ($\clModes$ modes), together with that of
the Poisson bracket approximation, discussed in Sec.~\ref{sec:pbme}, in which
the raising part of the force survives.
Part of this cancellation is known from the physical sector.  Evaluated
between physical states, the force and excess contributions each produce a
traced term with coefficient $\pm\tfrac14$, and these
cancel;\cite{nassimi2010,kelly2012} Kelly \emph{et al.} proved, moreover, that
$i\cL_m$ commutes with the projector onto the physical
sector.\cite{kelly2012}  The cancellation derived here holds for every Hermite
function, and it is this that makes each $\cV_K$ invariant.  It does not by
itself show that the physical sector, a four-dimensional subspace of the
eleven-dimensional even part of $\cV_2$ for $N=2$, is invariant; that is the
result of Ref.~\onlinecite{kelly2012}, which in the present variables takes
the following form.
\subsection{The physical sector and the subsystem QCLE}
\label{sec:physical}
A physical density corresponds to $\cc=\mathsf B\,{\rm vec}(\rho_W)$, where the
constant $11\times4$ matrix $\mathsf B$ (for $N=2$) contains the Hermite
coefficients of the polynomials $\Pi_{\lambda\lambda'}$.  Its left inverse in
the Hermite metric $\mathsf W={\rm diag}(\n!)$,
$\mathsf D=(\mathsf B^\dagger\mathsf W\mathsf B)^{-1}\mathsf B^\dagger\mathsf W$,
reads out the density matrix; applied to a vector with components outside
the physical sector, it returns the density matrix of the orthogonal
projection onto the sector in that metric.  Restricted to the
range of $\mathsf B$, Eq.~\eqref{eq:coeff} is the subsystem QCLE,
Eq.~\eqref{eq:qcle}; we have verified this identity in all matrix elements by
computer algebra (Sec.~S5).  Because the range of $\mathsf B$ is invariant,
Eq.~\eqref{eq:coeff} is, for physical initial data, an exact but redundant
form of Eq.~\eqref{eq:qcle}: the seven coefficient combinations outside the
physical sector remain zero.  How closely a numerical propagation keeps them
zero is measured by the relative residual
\begin{equation}
\begin{gathered}
\epsilon_{\rm phys}(t)=\frac{\|(\mathsf 1-\mathsf B\mathsf D)\,\cc\|_{\mathsf W}}
{\|\cc\|_{\mathsf W}},\\
\|\cc\|_{\mathsf W}^2=\int dR\,dP\,\sum_{\n}\n!\,|c_{\n}|^2 .
\end{gathered}
\label{eq:eps}
\end{equation}
For physical initial data, the graded representation contains no additional
dynamical information beyond Eq.~\eqref{eq:qcle} while the evolution remains
in the physical sector.  The Poisson-bracket approximation can drive the
density out of that sector.

\subsection{Structure of the Poisson bracket approximation}
\label{sec:pbme}
Because $i\cL_m^{\PB}=i\cL_m-i\cL_m'$, Eqs.~\eqref{eq:excess-raise} and
\eqref{eq:force-split} give the PBME in the same basis:
\begin{equation}
\partial_t\cc\big|_{\PB}=\Bigl[-\frac{P}{M}\,\partial_R+\mathsf R
+\mathsf G\,\partial_P\Bigr]\cc
+\frac14\sum_{ij}\partial_R\HH_{ij}\,\mathsf U^{ij}\,\partial_P\cc,
\label{eq:PBcoeff}
\end{equation}
with $\mathsf U^{ij}_{\m\n}=\delta_{\m,\n+\e_i+\e_j}$.  The approximation adds
to the exact dynamics precisely the raising part of the force, which acts on
the two components of a physical density, of order zero and two, in different
ways.
It raises the order-zero component, whose coefficient is $\Tr\rho_W$, into
order two.  With the traceless convention, this contribution lies in the
physical sector, so the PBME changes the evolution of the matrix readout
itself:
\begin{equation}
\partial_t\rho_W\big|_{\PB}-\partial_t\rho_W\big|_{\rm QCLE}
=\frac14\,\bigl(\partial_R\bh\bigr)\,\partial_P\Tr\rho_W .
\label{eq:inframe}
\end{equation}
This is the physical matrix element of the excess operator obtained by
Kelly \emph{et al.}:\cite{kelly2012} a force term, acting on the momentum
gradient of the total bath distribution $\Tr\rho_W$, that the exact back
reaction does not contain.  With the untraced convention the raised term
is not a physical density: it differs from the physical density of
$\tfrac14(\partial_Rh)\,\partial_P\Tr\rho_W$ by a multiple of the Gaussian
itself (Sec.~S4).
The PBME also raises the order-two components to order four, which is absent
from every physical density.  Define the PBME time-derivative amplitude
$q_{\PB}=-i\cL_m^{\PB}\rt^{\rm phys}/\varphi$, and let $\mathsf P_2$ be the
orthogonal projector onto $\cV_2$ in the Gaussian-weighted inner product.
The fourth-order part is
\begin{equation}
\begin{gathered}
(\mathsf 1-\mathsf P_2)\,q_{\PB}=\wick{q^{(4)}},\\
q^{(4)}=\frac14\sum_{ij}\partial_R\HH_{ij}\,z_iz_j
\sum_{\lambda\lambda'}\Pi^{(2)}_{\lambda\lambda'}(z)\,
\partial_P\rho_W^{\lambda'\lambda},
\end{gathered}
\label{eq:quartic}
\end{equation}
where $\Pi^{(2)}_{\lambda\lambda'}=\Pi_{\lambda\lambda'}+\delta_{\lambda\lambda'}$
is the homogeneous quadratic part of the projector polynomial, and
$\wick{\cdots}$ replaces each monomial $z^{\bm\alpha}$ by
$\He_{\bm\alpha}(z)$.  This Hermite-ordered expression is the component
orthogonal to $\cV_2$; it differs from the ordinary product by lower-order
terms.  For $N=2$, the defect occupies $\nQuarticModes$ of the $35$ Hermite
functions of fourth order.  Mapping symbols of subsystem operators are at
most quadratic, so fourth-order components are orthogonal to the
density-matrix readout when they are created.  They can affect the readout
later through the order-lowering part of the dynamics.  Under repeated
application, the PBME can increase the highest Hermite order by two at each
step, whereas the full mapping operator does not raise the order; physical
initial data therefore remain within $\cV_2$ under the full operator
(supplementary material, Sec.~S5).
The size of the fourth-order defect produced by the PBME acting on the exact
density defines a rate,
\begin{equation}
\mathcal D_4(t)=\frac{\bigl\|(\mathsf 1-\mathsf P_2)\,q_{\PB}(t)\bigr\|}
{\bigl\|\rt^{\,\rm phys}(t)/\varphi\bigr\|},
\label{eq:D4}
\end{equation}
where the function norm is the Gaussian-weighted $L^2$ norm integrated over
mapping variables and bath phase space:
$\|f\|^2=\int dR\,dP\,dz\,\varphi(z)|f(z,R,P)|^2$.
For $f=\sum_{\n} a_{\n}(R,P)\He_{\n}(z)$, this equals
$\int dR\,dP\sum_{\n}\n!|a_{\n}(R,P)|^2$, the coefficient norm in
Eq.~\eqref{eq:eps}.  The ratio has units of inverse time.  Its numerator is
the norm of a source linear in each of $\partial_R\bh$ and
$\partial_P\rho_W$; the pointwise source vanishes wherever either factor
vanishes.  Since $\mathcal D_4$ requires only a QCLE solution, it is a computable candidate indicator of PBME error, whose usefulness is tested in Sec.~\ref{sec:defect-results}.

\section{Model and computational methods}
\label{sec:methods}

\subsection{Model}

We use Tully's dual avoided crossing,\cite{tully1990}
$V_{11}=0$, $V_{22}=-Ae^{-BR^2}+E_0$, $V_{12}=V_{21}=Ce^{-DR^2}$, with
$A=0.10$, $B=0.28$, $C=0.015$, $D=0.06$, $E_0=0.05$ and $M=2000$ in atomic
units, and $V_b=0$ before the trace of $h$ is moved into it.  The initial
state is a minimum-uncertainty Gaussian wave packet on diabatic state $1$,
centred at $R_0=-15$ with mean momentum $P_0$ and width $\sigma_R=10/P_0$, so
that $\sigma_P=P_0/20$; $\rho_W(R,P,0)$ is its Wigner function.  We consider
$P_0=10$, $20$, $30$, $40$, $70$ and $100$ and propagate to $T=2M|R_0|/P_0$,
the time the packet centre needs to cross the interaction region at its
initial speed ($T=\TTPSeventy$ at $P_0=70$, rounded up to a multiple of the
time step $0.5$).  At $P_0=10$ the incident kinetic energy lies below the asymptotic
energy of state $2$, which is populated only while the packet crosses the
interaction region.  The observables are the diabatic populations
$\rho^{aa}(t)=\int dR\,dP\,\rho_W^{aa}$, the populations
$\int dR\,dP\,\langle k;R|\rho_W|k;R\rangle$ of the lower ($k=1$) and upper
($k=2$) adiabatic states $|k;R\rangle$, the eigenvectors of $h(R)$, and the
electronic coherence $\rho^{12}(t)=\int dR\,dP\,\rho_W^{12}$.  The error
measures defined below use $\rho^{22}$.

\subsection{Direct propagation of the coefficient equations}

For $N=2$ the physical initial data lie in the even part of $\cV_2$, so
Eq.~\eqref{eq:coeff} is propagated for $\nModesGH$ coefficient fields on a
uniform periodic $(R,P)$ grid with Fourier derivatives (Table~\ref{tab:params}).
The operator is applied as a whole, without splitting, and without forming a
matrix.  The time propagation uses the exponential of the semi-discrete
operator, applied to the state by the Arnoldi method\cite{saad1992,hochbruck1997}
with a subspace of $\ghMHigh$ vectors ($\ghMLow$ at $P_0=10$, where memory limits it).
We choose the substep from the endpoint Arnoldi residual,\cite{botchev2013}
using a relative residual tolerance of $\krylovTol$.  Because
$\mathsf B\,{\rm vec}(\rho_W)$ is real for a Hermitian $\rho_W$ and
$\mathsf R$ and $\mathsf G$ are real, the propagation is carried out in real
arithmetic.  The initial coefficients are $\mathsf B\,{\rm vec}(\rho_W)$, the
density matrix is read out with $\mathsf D$ at the recorded times, and the
physical sector is never imposed during the propagation.

To test the closure itself, the calculation at $P_0=\clPZero$ was repeated on a
$\clGrid$ grid in the basis of all even orders up to four ($\clModes$ modes, of
which $\clModesFour$ are of order four) and, for parity, in the basis of all
orders up to three ($\parModes$ modes, $\parModesOdd$ of odd order).  These
bases contain the sectors that the theory predicts to remain empty, so the
calculations would reveal a failure of the closure.

\subsection{Reference calculations}

\emph{Full quantum dynamics.}  The two-state wave packet is propagated with
the split-operator method\cite{feit1982} with a Fourier grid
representation\cite{kosloff1983} on $\qNRProdLow$ grid points
($\qNRProdHigh$ for $P_0\ge70$) with a time step of $\qDtProd$.  The
reference values are those of a calculation with a quarter of that step, twice the grid density and a larger box; the remaining
uncertainty, taken as the larger of the Richardson estimate at the finest step
and the changes produced by the grid and box refinements, is at most
$\qUncMax$ in $\rho^{22}(T)$.

\emph{Subsystem QCLE.}  Equation~\eqref{eq:qcle} is solved by two independent
codes.  The first advances the four real fields $\rho_W^{11}$,
$\rho_W^{22}$, ${\rm Re}\,\rho_W^{12}$ and ${\rm Im}\,\rho_W^{12}$ with Fourier
derivatives and the classical fourth-order Runge--Kutta method (RK4) at a
step below its stability limit ($\gridDtSmall$ or $\gridDtLarge$).  It uses the grid of the coefficient
propagation in every regime except $P_0=10$, where it was run on a grid
coarser by a factor of two in each direction.  Where this was tested
($P_0\ge20$), halving its step changes $\rho^{22}(T)$ by at most
$\gridDtHalfMax$ and halving its grid resolution in each direction by at most
$\gridSpatialMax$.  The second is a Strang splitting\cite{strang1968}
with the step $0.5$, in which the bath advection and the combined electronic
and force terms are each integrated exactly (Sec.~S6); it is second order in
the step.

\subsection{Poisson bracket mapping calculations}

The PBME density matrix is estimated without bias by sampling the
projected density of Eq.~\eqref{eq:rhophys}.  Mapping variables are drawn from
the importance density $\varphi|\Pi_{11}|/Z$ with $Z=4e^{-1/2}-1$, which keeps
the signed weights $Z\,{\rm sgn}\,\Pi_{11}$ bounded, and bath variables from
the initial Wigner function.  Each sample is propagated with Hamilton's
equations for $H_m$ at the step $0.5$, with a symmetric splitting in which the
mapping rotation at fixed $R$ is integrated exactly, and the density matrix is
read out with the mapping symbols of the subsystem operators.  The weights are
not renormalized, and a control variate built from the known initial density
matrix reduces the variance.  We use $\pbNSamples$ trajectories
($\pbNSamplesMin$ at $P_0=20$) in blocks of $\pbBlockSize$ and quote one
standard error of the mean over the blocks (Sec.~S6).  At $P_0=40$, halving the
time step changes $\rho^{22}(T)$ by $\pbDtChange$, a fraction $\pbDtOverSem$ of
the standard error.  For comparison we also report focused PBME,
using focused initial distribution sampling,\cite{bonella2003} with $\focNSamples$
trajectories, in which each mapping oscillator starts on a circle of fixed
action with a random phase.

\subsection{Comparison of the calculations}

We report the error of the QCLE with respect to full
quantum dynamics, $E_{\rm QCLE}=|\rho^{22}_{\rm QCLE}(T)-\rho^{22}_{\rm
quantum}(T)|$, with the QCLE value taken from the coefficient propagation, and
the change produced by the Poisson bracket approximation,
$E_{\rm PBME}=|\rho^{22}_{\rm PBME}(T)-\rho^{22}_{\rm QCLE}(T)|$, with the
PBME value taken from the projected-density estimator; both are formed from
unrounded values.  Differences between time series are formed only at instants
recorded by both calculations; no series is interpolated.

At $t=T$ the calculations are also compared on the densities themselves.
The partial Wigner transform of the converged wave function is evaluated on
the nodes of the coefficient propagation, and the four-field solution, whose
nodes lie at cell centres, is moved onto them by an exact Fourier
translation.  The PBME density is the projected-density estimator of the same
trajectories as above, resolved by phase-space cell, on cells of $\psCellRPThirty\times\psCellPPThirty$
in $R$ and $P$ at $P_0=30$ (Sec.~S7).  A grid density is compared with it
through its exact average over the same cells.  We report the negative volume
$N_a=\int dR\,dP\,\max(-\rho_W^{aa},0)$ of each diagonal density and the
distance $D_a=\int dP\,|n_a(P)-n_a^{\rm quantum}(P)|/\rho^{aa}_{\rm
quantum}(T)$ between momentum distributions $n_a(P)=\int dR\,\rho_W^{aa}$.

\subsection{Use of artificial intelligence tools}

A large language model, Anthropic's Claude, model Opus 5.5,  was used to help with writing the manuscript and checking the simulation codes. OpenAI's Codex assisted with editing, reference checks, formatting and submission preparation. The authors reviewed the output of these tools and take full responsibility for the
content.

\begin{table*}[t]
\caption{\label{tab:params}Parameters of the coefficient propagation: initial
width, final time, bath grid and box, number of exponential substeps and wall
time on one workstation.}
\begin{ruledtabular}
\begin{tabular}{rrrcccrr}
$P_0$ & $\sigma_R$ & $T$ & $n_R\times n_P$ & $R$ range & $P$ range & substeps & wall time (h)\\
\colrule
10  & $\sigRPTen$ & $\TTPTen$ & $\ghGridPTen$ & $\ghRboxPTen$ & $\ghPboxPTen$ & $\ghSubstepsPTen$ & $\ghHoursPTen$\\
20  & $\sigRPTwenty$ & $\TTPTwenty$ & $\ghGridPTwenty$ & $\ghRboxPTwenty$ & $\ghPboxPTwenty$ & $\ghSubstepsPTwenty$ & $\ghHoursPTwenty$\\
30  & $\sigRPThirty$ & $\TTPThirty$ & $\ghGridPThirty$ & $\ghRboxPThirty$ & $\ghPboxPThirty$ & $\ghSubstepsPThirty$ & $\ghHoursPThirty$\\
40  & $\sigRPForty$ & $\TTPForty$ & $\ghGridPForty$ & $\ghRboxPForty$ & $\ghPboxPForty$ & $\ghSubstepsPForty$ & $\ghHoursPForty$\\
70  & $\sigRPSeventy$ & $\TTPSeventy$ & $\ghGridPSeventy$ & $\ghRboxPSeventy$ & $\ghPboxPSeventy$ & $\ghSubstepsPSeventy$ & $\ghHoursPSeventy$\\
100 & $\sigRPOneHundred$ & $\TTPOneHundred$ & $\ghGridPOneHundred$ & $\ghRboxPOneHundred$ & $\ghPboxPOneHundred$ & $\ghSubstepsPOneHundred$ & $\ghHoursPOneHundred$\\
\end{tabular}
\end{ruledtabular}
\end{table*}

\section{Results}
\label{sec:results}

\subsection{The closure in computation}
\label{sec:closure-results}

In the basis of all even orders up to four, the calculation at $P_0=\clPZero$
was started from physical data and carried to $T=\clT$, through the
interaction region.  The fourth-order content
$E_4=\sum_{|\n|=4}\int dR\,dP\,|c_{\n}|^2$ was exactly zero at every recorded
time, as it must be when no element of the operator couples into that sector.
The sector is nevertheless dynamically connected to the rest: started instead
from fourth-order data with $E_4=\clCtlEFour$, the same calculation builds up
second-order content $E_2=\clCtlETwo$ from zero within $t=\clCtlT$.  In the basis containing both parities the odd
content likewise remained exactly zero, while odd initial data with
$E_{\rm odd}=\parCtlEOdd$ evolved to $\parCtlEOddFinal$ without generating
even content (Sec.~S7).  In the fourth-order calculation the residual of
Eq.~\eqref{eq:eps} stayed below $\clFrame$, so the physical sector is
invariant inside a space larger than $\cV_2$, as the projector result
requires.

\begin{figure*}[t]
\centering
\includegraphics[width=\textwidth]{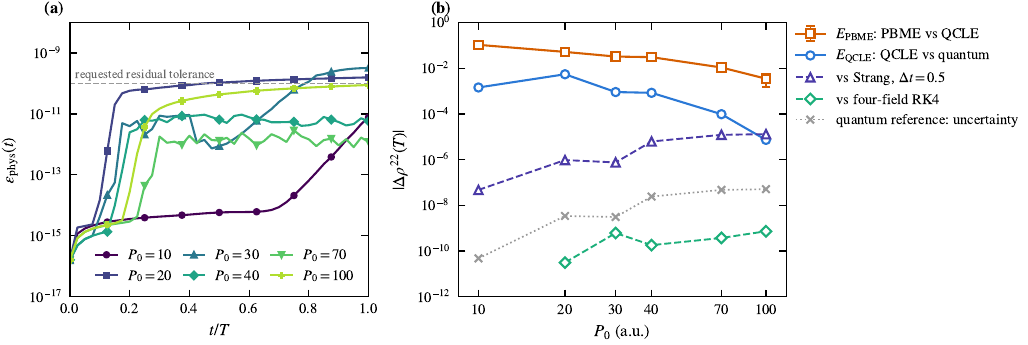}
\caption{\label{fig:numerics}(a) Physical-sector residual, Eq.~\eqref{eq:eps}, of
the coefficient propagation; the dashed line is the requested residual tolerance per
substep.  (b) The two physical errors, with two standard errors for
$E_{\rm PBME}$, and the differences of the coefficient propagation from the
Strang solution and, for $P_0\ge20$, the RK4 solution, against the incident
momentum; crosses: uncertainty of the quantum reference.}
\end{figure*}

In the production calculations the residual of the physical sector stays
below $\ghFrameMax$ in all six regimes [Fig.~\ref{fig:numerics}(a)].  It
rises as the packet enters the interaction region and remains at the level of
the residual tolerance requested per substep of the exponential propagator.  The
trace is conserved to $\ghTraceMax$, and the odd-parity content and the
Hermiticity error of the read-out density matrix are exactly zero, because
the operator contains no element that could produce them.

\subsection{Comparison with solvers of the subsystem QCLE}
\label{sec:benchmark}

On a common grid the coefficient propagation and the RK4 solution of
Eq.~\eqref{eq:qcle} agree in $\rho^{22}(T)$ to within $\eGridFinMax$, and to
within $\eGridMaxAll$ at every instant recorded by both
[Fig.~\ref{fig:numerics}(b); values for each regime in Sec.~S6].  The two
solve the same equation in different variables with unrelated time
integrators, so their agreement bounds the time-integration error of each; the
error of their common spatial discretization, measured by coarsening the grid
of the four-field solver, is below $\gridSpatialMax$.
At $P_0=10$ the four-field solver was run only on a grid coarser by two in
each direction.  Its difference from the coefficient propagation,
$\eGridPTen$, therefore compares solutions on different grids rather than the
two solvers; it is listed in Table~SIII but not shown in
Fig.~\ref{fig:numerics}(b), and at this momentum the independent check is the
Strang solution on the same grid.

The agreement extends to the densities themselves.  At $t=T$ the four-field
solution differs from the coefficient propagation, node by node, by at most
$\psGridPeakLowMax$ of the peak density at $P_0=20$ and $30$, and by
$\psGridPeakPSeventy$ and $\psGridPeakPOneHundred$ at $P_0=70$ and $100$.
At $P_0=\psRkList$ this difference is the error of the Runge--Kutta steps in
free flight.  Away from the crossing each Fourier component of the density is
advanced over a step by a known factor, which the scheme replaces by its
Taylor polynomial of degree four; the ratio of the two, accumulated over the
run and applied to the coefficient-propagation density, reproduces the
difference to within a relative residual of $\psRkResidMax$, and what remains is at
most $\psRkAfterMax$ of the peak (Sec.~S7).  For the diagonal densities the
ratio is one at zero wavenumber, so the populations do not register this
error.

The Strang solution differs from the coefficient propagation by amounts
ranging from $\eStrangPTen$ at $P_0=10$ to $\eStrangPOneHundred$ at $P_0=100$
[Fig.~\ref{fig:numerics}(b)].  The difference is the splitting error of the
scheme: halving the step reduces its difference from the RK4 solution by
factors between $\strangHalveMin$ and $\strangHalveMax$, as expected of a
second-order method (Sec.~S6).  At $P_0=100$ the
splitting error is $\strangOverQclePOneHundred$ times the error of the QCLE
itself, so a Strang calculation at this step cannot resolve the physical
error it would be used to measure; time-converged solutions are needed there.

The coefficient propagation carries eleven fields instead of four, and for
physical data seven combinations of them vanish identically; on the same grid
and machine it took between $\ghOverGridMin$ and $\ghOverGridMax$ times the
wall time of the four-field solver.  The costs are comparable, and we do not
propose the coefficient equations as a faster route to the subsystem QCLE;
their use lies elsewhere (Sec.~\ref{sec:discussion}).

\subsection{Quantum--classical error and Poisson bracket error}
\label{sec:errors}

Table~\ref{tab:results} and Fig.~\ref{fig:numerics}(b) separate the two
physical errors.  The final population of the QCLE differs from the quantum
one by $\eQclePTwenty$ at $P_0=20$, where the difference is largest, and by
$\eQclePOneHundred$ at $P_0=100$; beyond $P_0=20$ the difference decreases
steadily with the momentum.  The numerical uncertainty of the QCLE values, from
refinement of the time step and of the bath grid, is at most $\numUncHigh$
for $P_0\ge20$ and $\numUncPTen$ at $P_0=10$, where the slow packet makes the
calculation most demanding (Sec.~S6); the quantum reference is uncertain by at
most $\qUncMax$.  $E_{\rm QCLE}$ exceeds these by factors of at least
$\qcleOverNumMin$ and $\qcleOverQMin$, respectively, so it is a property of
the equation and not of its solution.  The final population varies
non-monotonically with $P_0$, reflecting the St\"uckelberg interference that the
model was designed to exhibit, and the QCLE follows this variation in every
regime.

\begin{table*}[t]
\caption{\label{tab:results}Final population of diabatic state $2$ and the two
physical errors.  QCLE: coefficient propagation; PBME: projected-density
estimator; focused: focused-sampling estimator.  Parentheses: one standard
error in units of the last digit.}
\begin{ruledtabular}
\begin{tabular}{rllllccr}
 & \multicolumn{4}{c}{$\rho^{22}(T)$} & \multicolumn{2}{c}{Errors} & \\
\cline{2-5}\cline{6-7}
$P_0$ & quantum & QCLE & PBME & focused & $E_{\rm QCLE}$ & $E_{\rm PBME}$ &
$E_{\rm PBME}/E_{\rm QCLE}$\\
\colrule
10  & $\qPopPTen$ & $\ghPopPTen$ & $\pbPopTabPTen$ & $\focPopTabPTen$ & $\eQcleTabPTen$ & $\ePbmeTabPTen$ & $\ratioPTen$\\
20  & $\qPopPTwenty$ & $\ghPopPTwenty$ & $\pbPopTabPTwenty$ & $\focPopTabPTwenty$ & $\eQcleTabPTwenty$ & $\ePbmeTabPTwenty$ & $\ratioPTwenty$\\
30  & $\qPopPThirty$ & $\ghPopPThirty$ & $\pbPopTabPThirty$ & $\focPopTabPThirty$ & $\eQcleTabPThirty$ & $\ePbmeTabPThirty$ & $\ratioPThirty$\\
40  & $\qPopPForty$ & $\ghPopPForty$ & $\pbPopTabPForty$ & $\focPopTabPForty$ & $\eQcleTabPForty$ & $\ePbmeTabPForty$ & $\ratioPForty$\\
70  & $\qPopPSeventy$ & $\ghPopPSeventy$ & $\pbPopTabPSeventy$ & $\focPopTabPSeventy$ & $\eQcleTabPSeventy$ & $\ePbmeTabPSeventy$ & $\ratioPSeventy$\\
100 & $\qPopPOneHundred$ & $\ghPopPOneHundred$ & $\pbPopTabPOneHundred$ & $\focPopTabPOneHundred$ & $\eQcleTabPOneHundred$ & $\ePbmeTabPOneHundred$ & $\ratioPOneHundred$\\
\end{tabular}
\end{ruledtabular}
\end{table*}

The change produced by the Poisson bracket approximation is larger in every
regime, by a factor of at least $\ratioMin$: $E_{\rm PBME}$ falls
from $\ePbmePTen$ at $P_0=10$ to $\ePbmePOneHundred$ at $P_0=100$ and is
resolved above its statistical uncertainty in all six regimes.  On this model,
therefore, it is the approximation to the mapping dynamics and not the
quantum--classical description that limits a PBME calculation.

\begin{figure*}[t]
\centering
\includegraphics[width=\textwidth]{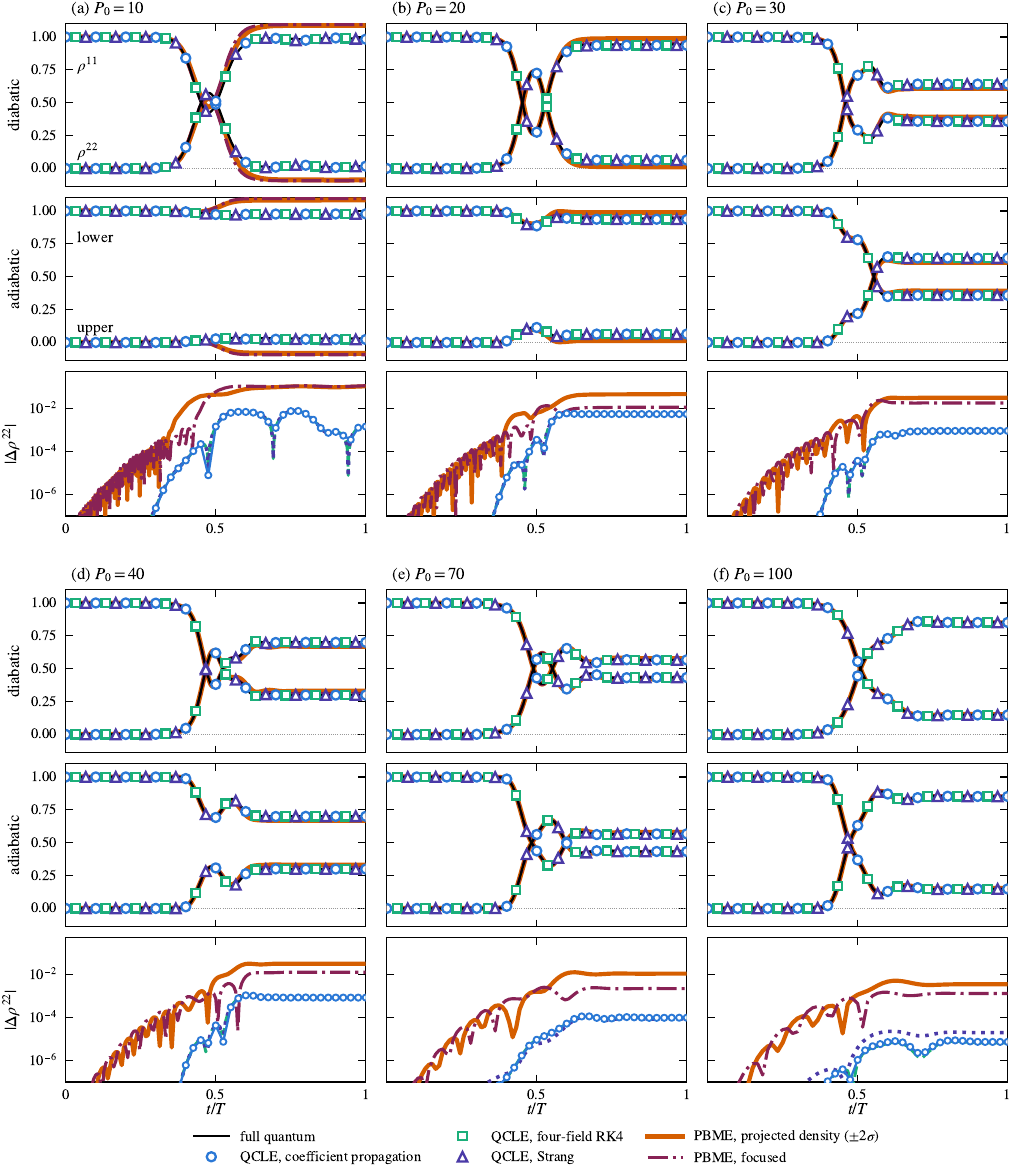}
\caption{\label{fig:populations}Populations against time for the six incident
momenta, from all calculations: diabatic states (upper panels), adiabatic
states (middle panels) and the deviation of $\rho^{22}$ from the quantum
result (lower panels).}
\end{figure*}

Figure~\ref{fig:populations} follows the diabatic and adiabatic populations
in time.  The three solutions of the subsystem QCLE coincide on the scale of
the figure and follow the quantum populations, whereas both PBME estimates
depart from them while the packet crosses the interaction region.  The lower
panels show the deviation of $\rho^{22}$ from the quantum result; every
calculation conserves the trace, so $|\Delta\rho^{11}|=|\Delta\rho^{22}|$.
The deviations arise during the crossing and, for $P_0\ge20$, change little
afterwards.  At every instant recorded by the quantum, QCLE and
projected-density PBME calculations at which the PBME deviation exceeds twice
its standard error, it is larger than that of the QCLE by at least a factor
$\ratioTimeMin$.

The PBME error is largest at the lowest momentum, where state $2$ is closed
asymptotically and its population rises while the packet crosses the
interaction region and falls back afterwards.  The PBME carries this
population below zero, to
$\rho^{22}(T)=\pbPopPTen\pm\pbSemPTen$, and the population of the upper
adiabatic state with it, to $\pbAdPopPTen\pm\pbAdSemPTen$.  Because the projected
estimator is unbiased for the PBME dynamics, the negative values belong to the
dynamics and not to the sampling.  The QCLE gives $\ghPopPTen$ and
$\ghAdPopPTen$ for the two quantities, against quantum values $\qPopPTen$ and
$\qAdPopPTen$.

Focused sampling gives a biased estimate of the PBME density.  Its final populations differ from
those of the projected-density estimator by between $\focDiffMin$ and
$\focDiffMax$, well above the statistical error of the latter, and lie closer
to the quantum result in \focCloserCount{} of the six regimes
(Table~\ref{tab:results}).  Its sampling and readout approximations can
therefore partly compensate the error of the Poisson bracket dynamics, so
agreement of such a calculation with full quantum results does not by itself
establish the accuracy of the underlying equation.

\begin{figure*}[t]
\centering
\includegraphics[width=\textwidth]{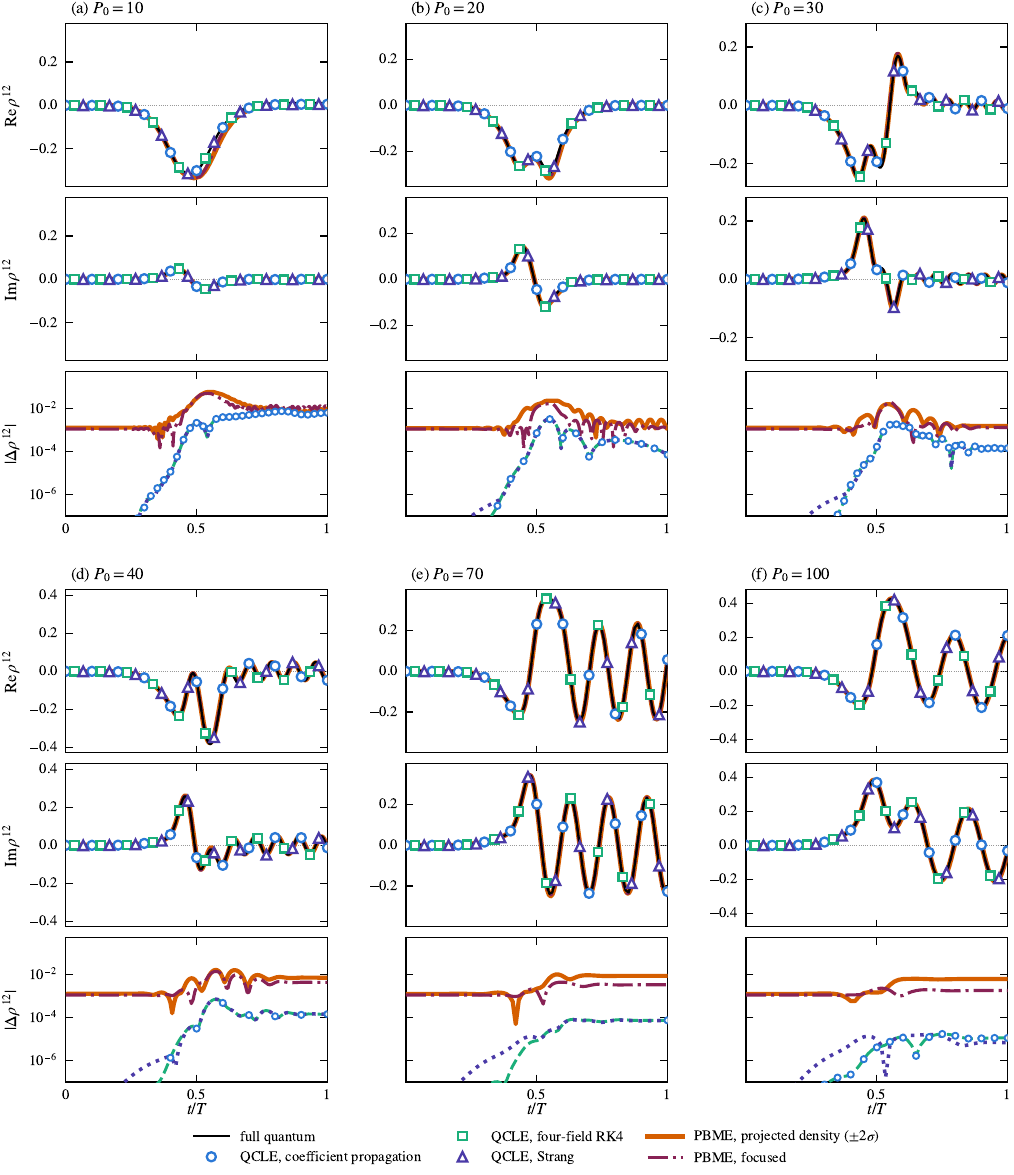}
\caption{\label{fig:coherence}Electronic coherence against time for the six
incident momenta, from all calculations: real part (upper panels), imaginary
part (middle panels) and modulus of the deviation from the quantum result
(lower panels).}
\end{figure*}

Figure~\ref{fig:coherence} shows the electronic coherence.  Integrated over
the bath, $\rho^{12}$ is the overlap of the components of the wave packet on
the two states.  It forms as the packet enters the interaction region, reaches
a modulus between $\cohPeakMin$ and $\cohPeakMax$ as the packet centre
crosses it, and afterwards persists as far as the two components continue to
overlap: at $P_0=70$ and $100$, where they move with nearly the same momentum,
its modulus at $T$ is still $\cohTPSeventy$ and $\cohTPOneHundred$, whereas at
$P_0=20$ it has fallen to $\cohTPTwenty$.  The three solutions of the
subsystem QCLE again coincide, the coefficient propagation and the RK4
solution to within $\cohGridMax$ on a common grid, and they differ from the
quantum coherence by at most $\cohQclePTen$ at $P_0=10$ and
$\cohQclePOneHundred$ at $P_0=100$.  While the coherence is still negligible,
the projected-density PBME estimate scatters about zero at the level of its
standard error, between $\cohSemMin$ and $\cohSemMax$.  After the crossing its
deviation from the quantum coherence reaches $\cohPbmePTen$ at $P_0=10$ and
$\cohPbmePOneHundred$ at $P_0=100$, at least $\cohPbmeOverSemMin$ standard
errors and between $\cohRatioMin$ and $\cohRatioMax$ times the largest
deviation of the QCLE in the same regime.  The largest deviation of the
focused estimate ranges from $\cohFocMin$ to $\cohFocMax$ over the six regimes
and again exceeds that of the QCLE in each.  The coherences therefore rank the two
approximations as the populations do.

\subsection{The errors in phase space}
\label{sec:phasespace}

\begin{figure*}[t]
\centering
\includegraphics[width=\textwidth]{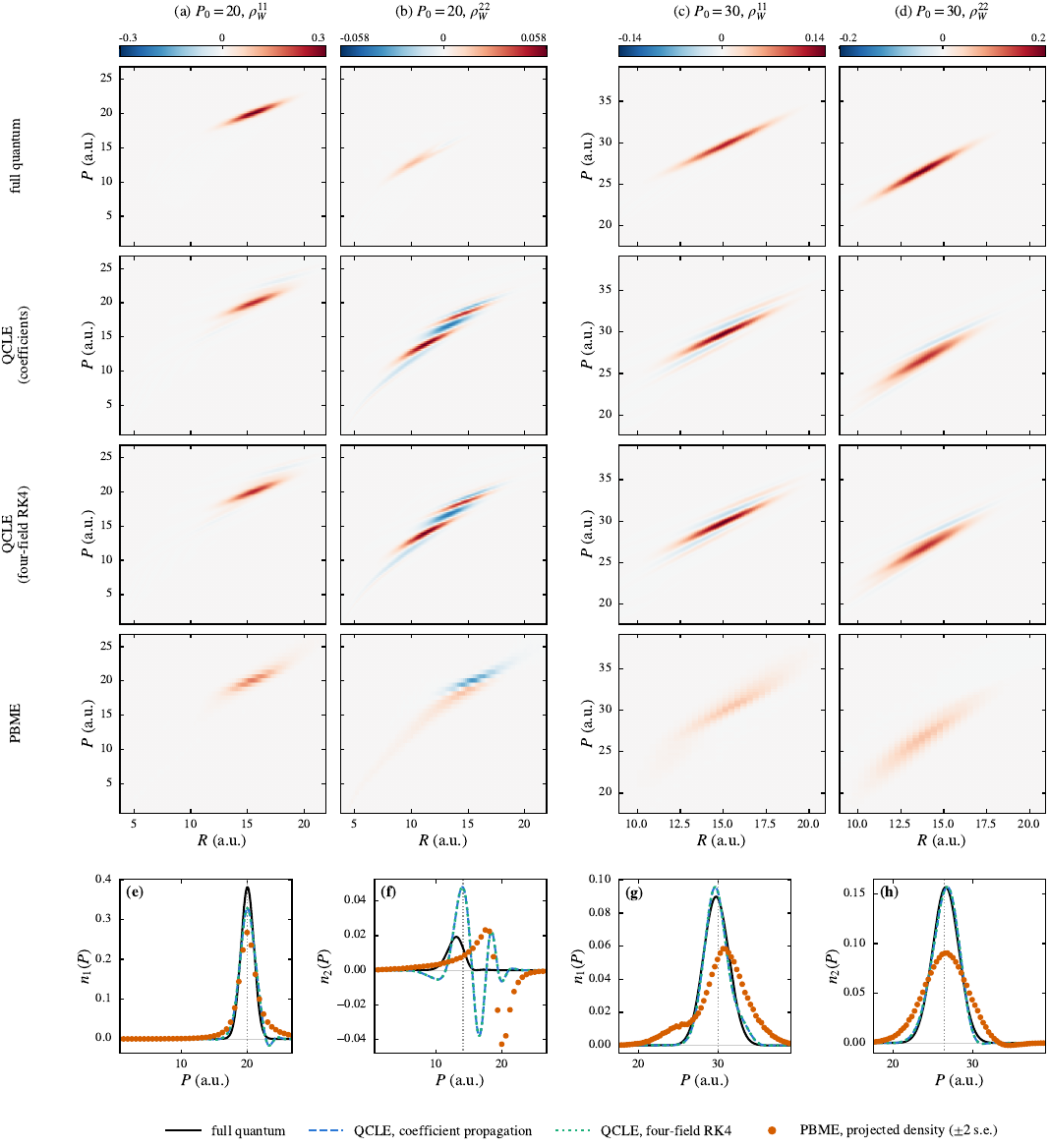}
\caption{\label{fig:phasespace}Diagonal densities at $t=T$ for $P_0=20$ and $30$.
(a)--(d) Rows from the top: full quantum dynamics, QCLE from the coefficient
propagation, QCLE from the four-field RK4 solver, and PBME on phase-space
cells.  (e)--(h) Momentum distributions of the column above; dotted lines
mark the asymptotic momenta of the two surfaces.}
\end{figure*}

The populations are integrals over phase space, and their agreement does not
imply that the densities agree.  Figure~\ref{fig:phasespace} compares the
diagonal densities at $t=T$ for $P_0=20$ and $30$; the other regimes are
shown in Sec.~S7, and $P_0=10$ is omitted because its terminal fields are not
resolved on the grid.  The colour scale of each column is set by the larger of
the quantum and QCLE maxima, and the PBME momentum distributions are shown
with two standard errors.  In the exact result each diagonal density is a single
packet, sheared by free flight.  On state $2$, whose asymptotic energy is
higher by $E_0$, a component of incident momentum $p$ leaves with the momentum
$(p^2-2ME_0)^{1/2}$; at the incident momentum $P_0$ the loss is
$\Delta P=P_0-(P_0^2-2ME_0)^{1/2}$, and in units of the momentum width
$\Delta P/\sigma_P$ is $\psShiftPTwenty$, $\psShiftPThirty$, $\psShiftPSeventy$
and $\psShiftPOneHundred$ at $P_0=20$, $30$, $70$ and $100$.  The momentum
dependence of the transition probability also reshapes both distributions, so
their centres need not lie at the asymptotic momenta marked in the lower
panels.  The exact diagonal densities are non-negative to within a negative
volume of $\psNegQMax$.

The second and third rows of Fig.~\ref{fig:phasespace} show the QCLE density
from the coefficient propagation and from the four-field RK4 solver.  They
cannot be told apart: in these two regimes they differ by at most
$\psGridPeakLowMax$ of the peak density (Sec.~\ref{sec:benchmark}), and in
all four regimes their negative volumes and momentum distances, discussed
next, agree to within $\psGridSameMeasures$.  The features described below
are therefore properties of the QCLE, not of the method used to solve it.

The QCLE places the two packets correctly, but its densities change sign in
bands alongside them, and the bands cancel in the populations.  At $P_0=20$
the bands of state $2$ fill the range between the momenta of the two surfaces: this density has a negative volume
$N_2=\psNegQcleTwoPTwenty$, against a population of $\ghPopPTwenty$, and its
momentum distribution changes sign and falls to $\psMomMinQcleTwoPTwenty$,
whereas the exact distribution is positive everywhere, with a peak of
$\psMomPeakQTwoPTwenty$ [Fig.~\ref{fig:phasespace}(f)].  The coherence, which
the QCLE carries along the mean of the two diabatic forces, extends over the
whole range of momenta between the two packets, whereas the exact coherence
is an interference pattern confined midway between them (Sec.~S7).  The
momentum distributions approach the exact ones as $\Delta P/\sigma_P$
decreases: the distance $D_2$ of the QCLE distribution
from the exact one falls from $\psMomQcleTwoPTwenty$ at $P_0=20$ to
$\psMomQcleTwoPThirty$ at $P_0=30$ and $\psMomQcleTwoPOneHundred$ at
$P_0=100$, and $D_1$ stays below $\psMomQcleOneMax$ (Table~SV).  The negative
regions of the densities persist, with volumes $N_1$ between
$\psNegQcleOneMin$ and $\psNegQcleOneMax$ in all four regimes (Table~SIV),
but they cancel almost entirely in the momentum distributions.

The PBME densities are distorted differently [bottom row of
Fig.~\ref{fig:phasespace}].  At $P_0=30$ the momentum distributions of both
states are broader than the exact ones, by factors of
$\psWidthPbmeOnePThirty$ and $\psWidthPbmeTwoPThirty$ in their standard
deviations [Fig.~\ref{fig:phasespace}(g),(h)].  From $P_0=30$ upward the PBME
also separates the two states in momentum by more than the exact dynamics:
the mean momenta of states $1$ and $2$ differ by $\psSplitPbmePThirty$,
$\psSplitPbmePSeventy$ and $\psSplitPbmePOneHundred$ at $P_0=30$, $70$ and
$100$, against $\psSplitQPThirty$, $\psSplitQPSeventy$ and
$\psSplitQPOneHundred$ exactly and $\psSplitQclePThirty$, $\psSplitQclePSeventy$
and $\psSplitQclePOneHundred$ in the QCLE.  At $P_0=20$ the PBME density of
state $2$ is negative near the momentum of state $1$ and positive over a broad
range below it [Fig.~\ref{fig:phasespace}(f)].  In every regime the PBME
momentum distributions lie further from the exact ones than those of the QCLE,
with $D_a$ larger by factors between $\psMomRatioMin$ and $\psMomRatioMax$ and
each at least $\psMomOverFloorMin$ times its statistical floor (Table~SV).
These distortions are consistent with the structure of the PBME force, which
combines the two diabatic forces in proportions set by the mapping variables
of each trajectory; over the physical initial density these proportions are
not confined between zero and one.

Over the whole density the comparison is less one-sided.  The relative $L_2$
distance of the cell-averaged QCLE density from the exact one exceeds that of
the PBME for $\rho_W^{22}$ at $P_0=\psLtwoQcleWorseTwoList$ and for the
coherence at $P_0=\psLtwoQcleWorseCohList$ (Table~SVI), because the bands of the QCLE, which cancel in the momentum
distributions, enter this distance in full.  The ordering of the
two errors found in the populations therefore holds for the momentum
distributions but not point by point, and at $P_0=20$, where $\Delta P$ is
several times $\sigma_P$, neither description reproduces the momentum
distribution of the transferred population, although the QCLE gives its
weight with an error of $\eQclePTwenty$, $\ratioPTwenty$ times smaller than
the change produced by the Poisson bracket approximation.

\subsection{The fourth-order defect as an error indicator}
\label{sec:defect-results}

\begin{figure*}[t]
\centering
\includegraphics[width=\textwidth]{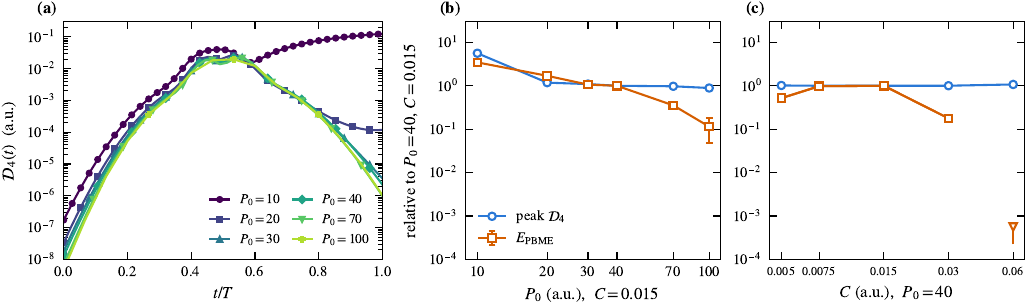}
\caption{\label{fig:defect}(a) Fourth-order defect rate, Eq.~\eqref{eq:D4}.  (b),
(c) Peak of $\mathcal D_4$ and $E_{\rm PBME}$ relative to their values at
$P_0=40$, $C=0.015$, along the momentum series and along a series of
diabatic couplings at $P_0=40$; the triangle is an upper bound on an
unresolved $E_{\rm PBME}$.}
\end{figure*}

The rate of Eq.~\eqref{eq:D4} requires only a solution of the QCLE and its
momentum gradient, which a grid calculation supplies.  We evaluate it on
Strang solutions with the step $\dhDt$ ($\dhDtPSeventy$ at $P_0=70$) on the
grids of Table~\ref{tab:params}; by the second-order scaling of
Sec.~\ref{sec:benchmark}, their populations differ from the converged ones by
less than $\dhSplitErr$, which is immaterial here.
Figure~\ref{fig:defect}(a) shows the rate against time.  It rises by several
orders of magnitude as the packet enters the interaction region, peaks when
the packet centre crosses it, and decays afterwards; for $P_0\ge20$ the curves nearly coincide when time is measured in
units of $T$, with peak values between $\dhPeakRangeLo$ and
$\dhPeakRangeHi$.  At $P_0=10$, where the passage is slowest, it does not decay
but grows through the second half of the propagation, to $\dhPeakPTen$ at
$t=T$.

Whether the rate indicates the size of the PBME error can be tested along
two directions in parameter space [Fig.~\ref{fig:defect}(b),(c)].  Along the
momentum series with $P_0\ge20$ its peak changes by a factor $\dhRangeMom$
while $E_{\rm PBME}$ changes by a factor $\ePbmeRangeMom$; only the slowest
regime, where both are largest, stands apart.  Along a series of diabatic
couplings $C=\scanCList$ at $P_0=40$ the peak changes by a factor
$\scanPeakRange$, while the PBME error is not even monotonic: it reaches
$\scanRespMax$ near the model's own coupling and falls below $\scanBound$ at
$C=\scanBoundC$, where it is no longer resolved.  For the coupling series the
QCLE reference is the Strang solution, whose splitting error at this momentum,
$\strangScanErr$, lies far below the resolved responses.  The fourth-order rate is thus
not a quantitative indicator of the PBME error on this model.  The reason is
visible in Eq.~\eqref{eq:quartic}: the rate is bilinear in $\partial_R\bh$ and
$\partial_P\rho_W$, and for this model $\partial_R\bh$ is dominated by the
diagonal, state-splitting term, so that its size changes by only a factor
$\scanGradSpan$ across the coupling series, whereas the coupling controls the
energy gap and with it the nonadiabatic dynamics that determine the error.  The rate also omits the part of the error generated
within the physical sector, Eq.~\eqref{eq:inframe}.  Finally, it requires
$\partial_P\rho_W$, which a trajectory calculation does not supply: estimated
from a PBME ensemble by kernel density differentiation, the relative error of
$\partial_P\rho_W$ decreases only slowly with the number of trajectories
(Sec.~S8).

\section{Discussion}
\label{sec:discussion}

In the symmetrized Moyal expansion that produces Eq.~\eqref{eq:mqcle-excess}, the
back-reaction force enters at leading order.  Because the force depends
quadratically on the mapping variables, it multiplies the mapping density by a
function that raises its Hermite order.  The excess coupling term is the next
order of the same expansion, and in the Gaussian--Hermite basis it is exactly
the raising operator that undoes this.  The exact quantum--classical dynamics
of the mapping oscillators therefore never generates structure of higher order
than it starts with, whereas the Poisson bracket, the leading classical part
of the expansion, lacks the counter-term.  The physical sector, whose
invariance Kelly \emph{et al.} established,\cite{kelly2012} thus lies within a
nested sequence of invariant spaces, one for every Hermite order, and the
Poisson bracket approximation breaks every one of them.

On the model studied, the Poisson bracket approximation introduces a larger
error than the quantum--classical description does, in the populations and
the coherences of every regime (Sec.~\ref{sec:errors}), and the comparison in
phase space refines this conclusion.  How well the QCLE represents momentum-resolved observables
depends on the momentum change between the surfaces relative to the momentum
width of the packet.  Where the change is below half of the width, the
momentum distributions of the QCLE differ from the exact ones by at most
$\psMomQcleHighMax$ (at $P_0=\psShiftHighList$).  At $P_0=30$ the distances
are $\psMomQcleOnePThirty$ for state $1$ and $\psMomQcleTwoPThirty$ for state
$2$, and at $P_0=20$ the momentum distribution of the transferred population
is qualitatively wrong, although the error in the population itself is
$\eQclePTwenty$.  The Poisson bracket approximation distorts the momentum
distributions of both states in every regime, including those in which its
population error is small.  Comparisons of mapping methods
on populations alone can therefore conceal errors of either kind.

For exact dynamics from physical initial data the coefficient equations
contain nothing that the subsystem QCLE does not, and they cost more; their
use is for dynamics or data that leave the physical sector.  The PBME is one
example, and Eq.~\eqref{eq:PBcoeff} writes it as an explicit Hermite
hierarchy, which makes it possible to settle a question that the present
calculations leave open: truncating Eq.~\eqref{eq:PBcoeff} at
second order retains exactly the part of the PBME error generated within the
physical sector, Eq.~\eqref{eq:inframe}, and removes the leak into higher
orders, so that comparing the truncated hierarchy with the full PBME would
apportion the error of Table~\ref{tab:results} between the two mechanisms.
Initial distributions used by trajectory methods that do not lie in the
physical sector, and approximations to $i\cL_m$ other than $i\cL_m^{\PB}$,
can be treated in the same way, since $i\cL_m$ evolves any density in $\cV_K$
within $\cV_K$.

The Hermite structure also limits what a change of mapping coordinates can
achieve.  The grading is defined by the isotropic Gaussian, and the linear maps
of the mapping space that preserve it and the Poisson bracket are the unitary
changes of the subsystem basis; $i\cL_m$ and $i\cL_m^{\PB}$ transform
covariantly under them, so no such choice removes the
defect, and nonlinear maps do not preserve the grading at all (Sec.~S9).
Within the linearized mapping family the defect is therefore a property of
the dynamics, not of the coordinates in which it is written.

The calculations concern one model with one bath coordinate and two states,
and the relative size of the quantum--classical and Poisson bracket errors,
and the failure of the fourth-order rate as an indicator, are established for
that model only.  The closure itself holds for any number of states and bath
coordinates.  The number of coefficient fields at second order grows as
$1+N(2N+1)$, against $N^2$ for the subsystem QCLE, and the bath is
discretized on a grid, so the coefficient equations are a tool for analysis
and for low-dimensional benchmarks, not for many-mode baths.

\section{Conclusion}

In a basis of Gaussian-weighted Hermite functions of the mapping variables,
the excess coupling term of the mapping quantum--classical Liouville equation
is a pure raising operator that cancels the raising part of the back-reaction
force.  The full mapping Liouville operator therefore preserves every finite
Hermite space,
and the mapping dynamics closes on a finite, sparse set of coefficient
equations that reduces to the subsystem QCLE on the physical sector.  The
Poisson bracket approximation retains the raising part of the force: it alters
the back reaction within the physical sector and lifts the density out of it,
into fourth order and beyond, with a fourth-order defect that has a closed
form.

For Tully's dual avoided crossing, direct propagation of the coefficient
equations confirms the closure and agrees with a converged independent solver
of the subsystem QCLE in the populations, the coherences and the phase-space
densities, to the accuracy of the solvers; a second-order splitting at a customary step, by
contrast, can have an error comparable to the physical effect it is meant to
resolve.  Against full quantum dynamics, the error introduced by the Poisson
bracket approximation exceeds that of the quantum--classical description in
every regime, by factors between $\ratioMin$ and $\ratioMax$ in the final
populations and between $\cohRatioMin$ and $\cohRatioMax$ in the largest
deviations of the coherences, and the size of the fourth-order defect does not
track it.  Resolved in phase space, the
Poisson bracket approximation also distorts the momentum distributions of both
states, broadening them or displacing them against each other, more than the
quantum--classical description does in every regime compared; at low incident
momentum neither reproduces the momentum distribution of the transferred
population.

\section*{Supplementary material}

See the supplementary material for the normalization of the physical mapping
density; the derivation of Eq.~\eqref{eq:mqcle-excess} from the Moyal product; the
Hermite algebra behind Eqs.~\eqref{eq:raise}--\eqref{eq:RG}; the
Poisson bracket approximation in both trace conventions; the computer-algebra
verification; the numerical methods, their convergence and the statistics of
the PBME estimators; the closure tests and the phase-space densities;
the estimation of the defect rate from trajectories; and changes of mapping
coordinates.

\begin{acknowledgments}
We thank the late Jeremy Schofield for his guidance throughout this work, and
Raymond Kapral for discussions of the quantum--classical Liouville equation
and its mapping representation.  The authors received no specific funding for
this work.
\end{acknowledgments}

\section*{Author declarations}
\subsection*{Conflict of interest}
The authors have no conflicts to disclose.

\subsection*{Author contributions}

\textbf{Sahand Nikzat:} Conceptualization (lead); data curation (lead);
formal analysis (lead); investigation (lead); methodology (lead); software
(lead); validation (lead); visualization (lead); writing -- original draft
(lead); writing -- review and editing (lead).

\begin{samepage}
\noindent\textbf{Ali Nassimi:}
Formal analysis (supporting); validation (supporting); writing -- review and
editing (supporting).
\end{samepage}

\begin{samepage}
\noindent\textbf{Rony Modak Parai:}
Formal analysis (supporting); validation (supporting); writing -- review and
editing (supporting).
\end{samepage}

\section*{Data availability}

The data that support the findings of this study are openly available in
GitHub at \url{https://github.com/sahandgit/mqcle-hermite-closure}.  The
repository contains the source code, the configuration files and the
computer-algebra scripts that reproduce every symbolic and numerical
statement reported here, together with the processed output of every
calculation.  Field archives larger than the repository's file-size limit
are available from the corresponding author upon reasonable request.

\bibliography{references}

\end{document}


\title{Supplementary material for ``Exact Gaussian--Hermite closure of the
mapping quantum--classical Liouville equation''}

\author{Sahand Nikzat}
\email{sahand.nikzat@mail.utoronto.ca}
\thanks{ORCID: \url{https://orcid.org/0009-0000-4224-4482}}
\affiliation{Chemical Physics Theory Group, Department of Chemistry,
University of Toronto, 80 St.\ George Street, Toronto, Ontario M5S 3H6,
Canada}
\author{Ali Nassimi}
\email{anasimi@gmail.com}
\thanks{ORCID: \url{https://orcid.org/0000-0002-2639-0962}}
\affiliation{Independent Researcher, St. John's, Newfoundland and Labrador,
Canada}
\author{Rony Modak Parai}
\email{ronyparai176@gmail.com}
\thanks{ORCID: \url{https://orcid.org/0009-0000-9785-6216}}
\affiliation{Chemical Physics Theory Group, Department of Chemistry,
University of Toronto, 80 St.\ George Street, Toronto, Ontario M5S 3H6,
Canada}

\date{October 11, 2026}

\maketitle

This supplement gives the derivations and numerical details behind
the results in the main text.  Its equations, figures and tables carry the
prefix S; references without this prefix point to the main text.

We follow the notation introduced in the main text.  The mapping coordinates
and momenta form the vector $x=(r_1,\dots,r_N,p_1,\dots,p_N)$, which we
rescale as $z=(2/\hbar)^{1/2}x$.  In these variables, the Gaussian weight
used in the Hermite expansion is $\varphi(z)=(2\pi)^{-N}e^{-z\cdot z/2}$.
The matrix $\HH$ is the real $2N\times2N$ representation of $h$, and
$\JJ=\bigl(\begin{smallmatrix}0&\one\\-\one&0\end{smallmatrix}\bigr)$ is
the symplectic matrix.  The multi-index $\n$ specifies the Hermite order
in each of the $2N$ mapping variables.

\section{Normalization of the physical mapping density}
\label{sec:S1}

The algebra of the physical mapping density is most transparent in the
complex amplitudes
\begin{equation}
w_\lambda=\frac{z_\lambda+i\,z_{N+\lambda}}{\sqrt2},\qquad \lambda=1,\dots,N .
\label{eq:w}
\end{equation}
The projector polynomial of \Meq{eq:rhophys} and the Weyl symbol of
$\hat a^\dagger_\lambda\hat a_{\lambda'}$ are
\begin{equation}
\begin{gathered}
\Pi_{\lambda\lambda'}=2\,w_\lambda w_{\lambda'}^*-\delta_{\lambda\lambda'},\\
\bigl(\hat a^\dagger_\lambda\hat a_{\lambda'}\bigr)_W
=\frac{1}{2\hbar}\Bigl[r_\lambda r_{\lambda'}+p_\lambda p_{\lambda'}
+i\bigl(r_\lambda p_{\lambda'}-p_\lambda r_{\lambda'}\bigr)
-\hbar\,\delta_{\lambda\lambda'}\Bigr]
=\frac12\bigl(w_\lambda^*w_{\lambda'}-\delta_{\lambda\lambda'}\bigr).
\end{gathered}
\label{eq:Pi-w}
\end{equation}
Under the Gaussian $\phi(x)\,dx=\varphi(z)\,dz$ the $w_\lambda$ are
independent complex Gaussian variables with
$\langle w_\lambda w_{\lambda'}^*\rangle=\delta_{\lambda\lambda'}$ and
$\langle w_\lambda w_{\lambda'}\rangle=0$, and Wick's theorem gives
\begin{equation}
\langle\Pi_{\lambda\lambda'}\rangle=\delta_{\lambda\lambda'},\qquad
\bigl\langle(\hat a^\dagger_k\hat a_{k'})_W\,\Pi_{\lambda\lambda'}\bigr\rangle
=\delta_{k\lambda}\,\delta_{k'\lambda'} .
\label{eq:wick-norm}
\end{equation}
Hence $\int dx\,\rho^{\rm phys}=\Tr\rho_W$ and
$\int dx\,(\hat a^\dagger_k\hat a_{k'})_W\,\rho^{\rm phys}=\rho_W^{k'k}
=\Tr\bigl(|k\rangle\langle k'|\,\rho_W\bigr)$.  By linearity the phase-space
average of the mapping symbol $O_m=\sum_{kk'}O_{kk'}(\hat
a^\dagger_k\hat a_{k'})_W$ of any subsystem operator $\hat O$ is
$\Tr(\hat O\rho_W)$.  With $h$ in place of $O$ the same symbol is the
mapping part of \Meq{eq:Hm},
$(2\hbar)^{-1}\bigl[x^{\mathsf T}\HH x-\hbar\Tr h\bigr]
=\sum_{\lambda\lambda'}h_{\lambda\lambda'}(\hat a^\dagger_\lambda\hat
a_{\lambda'})_W$, for every Hermitian $h$.

Equation~\eqref{eq:wick-norm} fixes three conventions at once: the prefactor
$2/\hbar$ of $\Pi_{\lambda\lambda'}$, its constant term, and the pairing of
$\Pi_{\lambda\lambda'}$ with $\rho_W^{\lambda'\lambda}$.  The opposite pairing
leaves the diagonal entries and real parts of the coherences unchanged,
but reverses the signs of their imaginary parts.

Two consequences are used in the main text.  First, since
$w_\lambda^*w_{\lambda'}-\delta_{\lambda\lambda'}$ is a sum of products of two
different variables and of $\tfrac12\He_2$ terms, every mapping symbol
$(\hat a^\dagger_k\hat a_{k'})_W$ is a pure second-order Hermite function.  The
readout of subsystem observables from any mapping density therefore involves
only its second-order Hermite coefficients; components of order zero, or of
order four and higher, are invisible to it at the instant they are present.
Second, the Hermite coefficients of the projector polynomial are
\begin{equation}
\begin{gathered}
\Pi_{\lambda\lambda}=1+\He_2(z_\lambda)+\He_2(z_{N+\lambda}),\\
\Pi_{\lambda\lambda'}=\He_{\e_\lambda+\e_{\lambda'}}
+\He_{\e_{N+\lambda}+\e_{N+\lambda'}}
-i\,\He_{\e_\lambda+\e_{N+\lambda'}}
+i\,\He_{\e_{\lambda'}+\e_{N+\lambda}}\qquad(\lambda\neq\lambda'),
\end{gathered}
\label{eq:Pi-hermite}
\end{equation}
which are the columns of the matrix $\mathsf B$ of
Sec.~\Mref{sec:physical}.  In particular the lowest coefficient of a physical
density is $c_{\bm 0}=\Tr\rho_W$.

\section{The mapping QCLE from the Moyal product}
\label{sec:S2}

Following Kim, Nassimi and Kapral,\cite{kim2008} each subsystem state is
represented by a singly excited mapping oscillator, $|\lambda\rangle\to\hat
a^\dagger_\lambda|0\rangle$, and each subsystem operator by
$\sum_{\lambda\lambda'}O_{\lambda\lambda'}\hat a^\dagger_\lambda\hat
a_{\lambda'}$.  On the singly excited subspace the representation preserves
products, because $\hat a^\dagger_\lambda\hat a_{\lambda'}\hat
a^\dagger_\mu\hat a_{\mu'}=\delta_{\lambda'\mu}\hat a^\dagger_\lambda\hat
a_{\mu'}+\hat a^\dagger_\lambda\hat a^\dagger_\mu\hat a_{\lambda'}\hat a_{\mu'}$
and the last term annihilates every singly excited state.  The QCLE,
\Meq{eq:qcle-op}, with the operator order kept in each product, therefore
holds unchanged for operators of the oscillators on that
subspace.\cite{kapral1999,kim2008}  A
Wigner transform over the oscillator coordinates replaces each operator by its
Weyl symbol and each product by the Moyal product
\begin{equation}
A\star B=A\,\exp\Bigl(\frac{i\hbar}{2}\Lambda\Bigr)B,\qquad
\Lambda=\sum_{ij}\overleftarrow{\partial}_{x_i}\,\JJ_{ij}\,\overrightarrow{\partial}_{x_j},
\label{eq:moyal}
\end{equation}
so that $A\Lambda B=\{A,B\}_x$ is the Poisson bracket in the mapping
variables.  The symbol of $\hat H_W$ is $H_m$ of \Meq{eq:Hm}.

The commutator part of $-i\hat\cL\hat\rho_W$ becomes
\begin{equation}
-\frac{i}{\hbar}\bigl(H_m\star\rho-\rho\star H_m\bigr)
=\frac{2}{\hbar}\,H_m\sin\Bigl(\frac{\hbar}{2}\Lambda\Bigr)\rho
=\{H_m,\rho\}_x ,
\label{eq:sine}
\end{equation}
exactly, because every term of the sine series beyond the first contains at
least three derivatives of $H_m$, which is quadratic in $x$.  With
$\partial_PH_m=P/M$, the bath term becomes
\begin{equation}
\frac12\bigl(\partial_RH_m\star\partial_P\rho+\partial_P\rho\star\partial_RH_m\bigr)
-\frac{P}{M}\,\partial_R\rho
=\partial_RH_m\cos\Bigl(\frac{\hbar}{2}\Lambda\Bigr)\partial_P\rho
-\frac{P}{M}\,\partial_R\rho .
\label{eq:cosine}
\end{equation}
The potential gradient $\partial_RH_m$ is also quadratic in $x$, with
$\partial_{x_i}\partial_{x_j}\partial_RH_m=\hbar^{-1}\partial_R\HH_{ij}$, so the
cosine ends after its second term:
\begin{equation}
\partial_RH_m\cos\Bigl(\frac{\hbar}{2}\Lambda\Bigr)\partial_P\rho
=\partial_RH_m\,\partial_P\rho
-\frac{\hbar}{8}\sum_{ij}\bigl(\JJ^{\mathsf T}\partial_R\HH\,\JJ\bigr)_{ij}\,
\frac{\partial^2}{\partial x_i\partial x_j}\,\partial_P\rho .
\label{eq:cosine2}
\end{equation}
The matrix $\HH$ is the real form of multiplication by the complex matrix $h$,
so it commutes with $\JJ$, which is the real form of multiplication by $-i$;
hence $\JJ^{\mathsf T}\partial_R\HH\,\JJ=\partial_R\HH$ and the last term is
$-i\cL_m'\rho$.  The remaining terms combine into
$\{H_m,\rho\}_{\cX}=-i\cL_m^{\PB}\rho$, and collecting
Eqs.~\eqref{eq:sine}--\eqref{eq:cosine2} gives
\Meqs{eq:mqcle-full}{eq:mqcle-excess}.  Nothing has been neglected: both series terminate because the
Hamiltonian is quadratic in the mapping variables.

Three remarks complete the derivation.  (i) For real symmetric $h$,
$\sum_{ij}\partial_R\HH_{ij}\partial_{x_i}\partial_{x_j}
=\sum_{\lambda\lambda'}\partial_Rh_{\lambda\lambda'}
(\partial_{r_\lambda}\partial_{r_{\lambda'}}+\partial_{p_\lambda}\partial_{p_{\lambda'}})$,
the operator of Refs.~\onlinecite{kim2008} and \onlinecite{kelly2012}.  For
$h=A+iB$ with $B\ne0$ the $\HH$ form contains in addition
$2\sum_{\lambda\lambda'}\partial_RB_{\lambda\lambda'}\,\partial_{p_\lambda}\partial_{r_{\lambda'}}$,
which the real-symmetric form omits; without it the subsystem QCLE is not
reproduced (Sec.~\ref{sec:S5}).  (ii) Dynamical variables evolve with the same
operator, $dB_m/dt=i\cL_mB_m$,\cite{kim2008,kelly2012} because $i\cL_m$ is
anti-self-adjoint in the phase-space inner product: $i\cL_m^{\PB}$ generates a
volume-preserving flow, and $i\cL_m'$ contains three derivatives with an
$R$-dependent coefficient.  (iii) Writing $h=\bh+c\,\one$ with $c=N^{-1}\Tr h$,
the mapping Hamiltonian with $h$ and $V_b$ and the one with $\bh$ and $V_b+c$
differ by
\begin{equation}
\delta H_m=c(R)\bigl[\one_m(x)-1\bigr],\qquad
\one_m(x)=\sum_\lambda(\hat a^\dagger_\lambda\hat a_\lambda)_W
=\frac{x\cdot x-N\hbar}{2\hbar},
\label{eq:trace-shift}
\end{equation}
and the two excess coupling terms by
$i\cL_m'(h)-i\cL_m'(\bh)=\tfrac{\hbar}{8}\partial_Rc\sum_i\partial_{x_i}^2\partial_P$.
The operator $\sum_\lambda\hat a^\dagger_\lambda\hat a_\lambda-1$ vanishes on
the singly excited subspace, but its symbol $\one_m-1$ is not zero, so the
Liouville operator of \Meq{eq:mqcle-full} changes.  The change vanishes on the
physical densities: the flow generated by $\one_m$ multiplies every
$w_\lambda$ by a common phase, which leaves $\varphi$ and every
$\Pi_{\lambda\lambda'}$ unchanged, and in the notation of Sec.~\ref{sec:S3}
the two force terms combine into
$\partial_Rc\,\bigl[\tfrac14\sum_i\cT_{ii}-1\bigr]\partial_P$, which annihilates
every physical amplitude.  Both conventions therefore reproduce the subsystem
QCLE, as the computer algebra confirms (Sec.~\ref{sec:S5}), but
$i\cL_m^{\PB}$ and $i\cL_m'$ change separately, and the PBME with them.
Using $h$ in one of $H_m$ and $i\cL_m'$ and $\bh$ in the other adds
$\pm\tfrac{\hbar}{8}\partial_Rc\sum_i\partial_{x_i}^2\partial_P$ to $i\cL_m$, a
pure raising operator in the Hermite basis, and destroys the closure; the
computer algebra confirms this for $h$ in $H_m$ and $\bh$ in $i\cL_m'$
(Sec.~\ref{sec:S5}).

\section{Hermite algebra and the coefficient equations}
\label{sec:S3}

\subsection{One-dimensional identities}

With $\varphi_1(z)=(2\pi)^{-1/2}e^{-z^2/2}$ the probabilists' Hermite
polynomials satisfy $\varphi_1\He_n=(-1)^n\,d^n\varphi_1/dz^n$, so that
\begin{equation}
\frac{d}{dz}\bigl(\varphi_1\He_n\bigr)=-\varphi_1\He_{n+1},\qquad
z\He_n=\He_{n+1}+n\He_{n-1},\qquad
\frac{d\He_n}{dz}=n\He_{n-1},
\label{eq:1d}
\end{equation}
and $\int dz\,\varphi_1\He_m\He_n=n!\,\delta_{mn}$.  For
$\varphi\He_{\n}=\prod_i\varphi_1(z_i)\He_{n_i}(z_i)$ all three hold in each
variable.  Two applications of the first give \Meq{eq:raise}; two of the
second give \Meq{eq:force-split}.

\subsection{The coupling of the force and the excess term}

For a density $f=\varphi q$ with amplitude $q$, $\partial_{z_i}f=\varphi(\partial_{z_i}-z_i)q$
and hence
\begin{equation}
\varphi^{-1}\,\partial_{z_i}\partial_{z_j}\bigl(\varphi q\bigr)
=\bigl[\partial_{z_i}\partial_{z_j}-z_i\partial_{z_j}-z_j\partial_{z_i}
+(z_iz_j-\delta_{ij})\bigr]q .
\label{eq:conjugation}
\end{equation}
In the variables $z$ the bath-momentum term differentiates the amplitude
with respect to $P$ and multiplies it by
$\partial_RV_b+\tfrac14\sum_{ij}\partial_R\HH_{ij}(z_iz_j-\delta_{ij})$.  The
excess term $-i\cL_m'$ takes the same derivative and applies
$-\tfrac14\sum_{ij}\partial_R\HH_{ij}\,\varphi^{-1}\partial_{z_i}\partial_{z_j}\varphi$.
By Eq.~\eqref{eq:conjugation} their sum is
\begin{equation}
\Bigl[\partial_RV_b+\frac14\sum_{ij}\partial_R\HH_{ij}\,\cT_{ij}\Bigr]\partial_Pq,
\qquad
\cT_{ij}=z_i\partial_{z_j}+z_j\partial_{z_i}-\partial_{z_i}\partial_{z_j} .
\label{eq:Tij}
\end{equation}
The multiplication by $z_iz_j$, the only operation that raises the polynomial
degree, has cancelled; in $\cT_{ij}$ every factor of $z$ is accompanied by a
derivative.  Its action on $\He_{\n}$ follows from Eq.~\eqref{eq:1d}:
\begin{equation}
z_i\partial_{z_j}\He_{\n}=n_j\He_{\n+\e_i-\e_j}
+n_j(n_i-\delta_{ij})\He_{\n-\e_i-\e_j},\qquad
\partial_{z_i}\partial_{z_j}\He_{\n}=n_j(n_i-\delta_{ij})\He_{\n-\e_i-\e_j},
\label{eq:Tparts}
\end{equation}
and adding the three pieces gives \Meq{eq:T}.  For $i=j$ it reduces to
$\cT_{ii}\He_{\n}=2n_i\He_{\n}+n_i(n_i-1)\He_{\n-2\e_i}$.

\subsection{The mapping flow}

The mapping part of the Poisson bracket is $\{H_m,f\}_x=-(Kx)\cdot\nabla_xf$
with $K=\hbar^{-1}\JJ\HH$.  Because $\HH$ is symmetric and commutes with
$\JJ$, $K^{\mathsf T}=-\hbar^{-1}\HH\JJ=-K$: the flow is a rotation, and it is
the same linear map in the variables $z$.  On $f=\varphi q$,
\begin{equation}
-\sum_{ij}K_{ij}z_j\partial_{z_i}(\varphi q)
=\varphi\Bigl[-\sum_{ij}K_{ij}z_j\partial_{z_i}q+\sum_{ij}K_{ij}z_iz_j\,q\Bigr],
\end{equation}
and the last sum vanishes because $K$ is antisymmetric.  By
Eq.~\eqref{eq:1d},
$-\sum_{ij}K_{ij}z_j\partial_{z_i}\He_{\n}=-\sum_{ij}K_{ij}n_i\bigl[\He_{\n-\e_i+\e_j}
+(n_j-\delta_{ij})\He_{\n-\e_i-\e_j}\bigr]$.  The coefficient $n_i(n_j-\delta_{ij})$ and
the function $\He_{\n-\e_i-\e_j}$ are symmetric under $i\leftrightarrow j$, so
the second term vanishes against the antisymmetric $K_{ij}$, and the rotation
acts as $\mathsf R=-\sum_{ij}K_{ij}\mathsf N^{ij}$ of \Meq{eq:RG}.  It
preserves the order of every Hermite function.

\subsection{The coefficient equations and the physical sector}

The bath advection $-(P/M)\partial_R$ does not act on $z$.  Collecting the
three parts gives \Meqs{eq:coeff}{eq:RG}.  Every element of $\mathsf R$ and
$\mathsf T^{ij}$ couples $\n$ to a multi-index of the same order or of order
lower by two, so the equations for the coefficients of order $k$ involve those
of orders $k$ and $k+2$ only.  Information flows down the Hermite ladder and
never up; the parity of the order is conserved; and every $\cV_K$ is invariant.
The coefficient equations have been verified against the Liouville operator
of \Meq{eq:mqcle-full} on every Hermite function of order up to four
(Sec.~\ref{sec:S5}).

For $N=2$ the even part of $\cV_2$ has $1+N(2N+1)=11$ coefficient fields, and
the $N^2=4$ columns of $\mathsf B$ are the coefficient vectors of
Eq.~\eqref{eq:Pi-hermite}.  The readout
$\mathsf D=(\mathsf B^\dagger\mathsf W\mathsf B)^{-1}\mathsf B^\dagger\mathsf W$
is the left inverse of $\mathsf B$ that is orthogonal in the Hermite metric
$\mathsf W={\rm diag}(\n!)$; the unweighted pseudoinverse is not, and does not
recover $\rho_W$ from a vector with components outside the physical sector.
For a vector inside it, $\mathsf D$ and the readout through mapping symbols of
Sec.~\ref{sec:S1} coincide.

\subsection{Mode ordering used by the code}

For $N=2$ the production code orders the eleven coefficient fields of the
even part of $\cV_2$ lexicographically in
$\n=(n_{r_1},n_{r_2},n_{p_1},n_{p_2})$, Table~\ref{tab:modes}.  Mode $0$
carries $c_{\bm 0}=\Tr\rho_W$; the other ten span the second-order block, in
which the four physical combinations of Eq.~\eqref{eq:Pi-hermite} lie.

\begin{table}
\caption{\label{tab:modes}Ordering of the eleven coefficient fields for $N=2$, with
$\n=(n_{r_1},n_{r_2},n_{p_1},n_{p_2})$.}
\begin{ruledtabular}
\begin{tabular}{rlrl}
0 & $(0,0,0,0)$ & 6 & $(0,2,0,0)$\\
1 & $(0,0,0,2)$ & 7 & $(1,0,0,1)$\\
2 & $(0,0,1,1)$ & 8 & $(1,0,1,0)$\\
3 & $(0,0,2,0)$ & 9 & $(1,1,0,0)$\\
4 & $(0,1,0,1)$ & 10 & $(2,0,0,0)$\\
5 & $(0,1,1,0)$ & & \\
\end{tabular}
\end{ruledtabular}
\par\smallskip{\footnotesize Alt text: a table listing the integer label and the multi-index of each of the
eleven modes.}
\end{table}

\section{The Poisson bracket approximation}
\label{sec:S4}

\subsection{Liouville operator}

Removing $i\cL_m'$ from $i\cL_m$ leaves uncancelled the raising part
of the force, which carries each component $c_{\n}\varphi\He_{\n}$ into
$\tfrac14\sum_{ij}\partial_R\HH_{ij}\,\partial_Pc_{\n}\,\varphi\He_{\n+\e_i+\e_j}$.
The PBME in \Meq{eq:PBcoeff} consists of the full coefficient operator
and this additional term, with
$\mathsf U^{ij}_{\m\n}=\delta_{\m,\n+\e_i+\e_j}$.  Acting on a physical density
it produces two terms.

\subsection{The raised trace}

The order-zero component $c_{\bm 0}=\Tr\rho_W$ is raised into
$\tfrac14\partial_P\Tr\rho_W\sum_{ij}\partial_R\HH_{ij}\He_{\e_i+\e_j}$.  For any
Hermitian matrix $a$ with real form $\mathbb A$, Eq.~\eqref{eq:Pi-w} gives
$z^{\mathsf T}\mathbb Az=2\sum_{\lambda\lambda'}a_{\lambda\lambda'}w_\lambda^*w_{\lambda'}
=\sum_{\lambda\lambda'}\Pi_{\lambda\lambda'}a_{\lambda'\lambda}+\Tr a$, and
$\Tr\mathbb A=2\Tr a$, so that
\begin{equation}
\sum_{ij}\mathbb A_{ij}\He_{\e_i+\e_j}
=\sum_{ij}\mathbb A_{ij}(z_iz_j-\delta_{ij})
=\sum_{\lambda\lambda'}\Pi_{\lambda\lambda'}\,a_{\lambda'\lambda}-\Tr a .
\label{eq:quadratic-identity}
\end{equation}
With $a=\partial_Rh$ the raised trace is
\begin{equation}
\frac14\,\partial_P\Tr\rho_W\Bigl[\sum_{\lambda\lambda'}\Pi_{\lambda\lambda'}
\bigl(\partial_Rh\bigr)_{\lambda'\lambda}-\Tr\partial_Rh\Bigr].
\label{eq:raised-trace}
\end{equation}
In the traceless convention $\Tr\partial_R\bh=0$ and the term is the physical
density whose density matrix is $\tfrac14(\partial_R\bh)\,\partial_P\Tr\rho_W$,
\Meq{eq:inframe}.  Kelly \emph{et al.} obtained this matrix element in
Eq.~(B8) of Ref.~\onlinecite{kelly2012}.  In the Hermite expansion it comes
from raising the order-zero component.  In the untraced convention the term is the physical density
of $\tfrac14(\partial_Rh)\,\partial_P\Tr\rho_W$ minus
$\tfrac14\Tr(\partial_Rh)\,\partial_P\Tr\rho_W$ times the Gaussian itself.  The
Gaussian alone is not a physical density: every physical density with a
nonzero trace carries second-order components with it,
Eq.~\eqref{eq:Pi-hermite}.  The Gaussian component is invisible to the readout
of observables (Sec.~\ref{sec:S1}) but is raised again by $i\cL_m^{\PB}$ at
later times.  The two conventions therefore give different Poisson bracket
approximations with different errors within the physical sector, which is why
the convention has to be stated with every PBME result.

\subsection{The fourth-order defect}

The second-order components of a physical amplitude are
$\sum_{\lambda\lambda'}\wick{\Pi^{(2)}_{\lambda\lambda'}}\rho_W^{\lambda'\lambda}$,
since $\Pi_{\lambda\lambda'}=\wick{\Pi^{(2)}_{\lambda\lambda'}}+\delta_{\lambda\lambda'}$
by Eq.~\eqref{eq:Pi-hermite}.  For $|\n|=2$ the raised function is
$\He_{\n+\e_i+\e_j}=\wick{z^{\n}z_iz_j}$, so the fourth-order part of the PBME
deviation is the Hermite-ordered product $\wick{q^{(4)}}$ of
\Meq{eq:quartic}.  The plain product $q^{(4)}$ differs from it by terms of
orders zero and two, which are not orthogonal to $\cV_2$; for $N=2$ its
monomial expansion has $\symTermsTop$ terms, that of $\wick{q^{(4)}}$
$\symTermsHerm$.  For generic $h$ and $\rho_W$ the defect occupies
$\nQuarticModes$ of the $\symOrderFourTotal$ Hermite functions of order four,
for real symmetric and complex Hermitian $h$ alike.  Each component is
bilinear in $\partial_R\HH$ and $\partial_P\rho_W$.  Under repeated
application of $i\cL_m^{\PB}$ the highest order reached from a physical
density is $2+2k$ after $k$ applications, verified for $k\le\symPowers$,
while under $i\cL_m$ it stays at two.

The defect rate of \Meq{eq:D4} uses the norm of \Meq{eq:eps}, which is the
norm of the amplitude in $L^2(\varphi\,dz\,dR\,dP)$.  Since $i\cL_m$
keeps a physical density in $\cV_2$, $(\mathsf 1-\mathsf P_2)q_{\PB}$ equals the
fourth-order part of the deviation, $\wick{q^{(4)}}$, and is evaluated from
the $\nQuarticModes$ Hermite coefficients of that polynomial, each a bilinear
form in $\partial_R\bh$ and $\partial_P\rho_W$ computed on the grid.

\section{Computer-algebra verification}
\label{sec:S5}

The statements of Sec.~\Mref{sec:theory} were checked by exact computer
algebra with arbitrary functions $h(R)$, $V_b(R)$ and $\rho_W(R,P)$, so each
check is a proof for the stated number of states and bath coordinates and not a
numerical test.  Densities are represented as the Gaussian times a polynomial
amplitude, the operators are built from their definitions (the Poisson bracket
in $(r,p,R,P)$ and $i\cL_m'$ with the coefficient $\hbar/8$), Hermite coefficients are
extracted by exact peeling of the highest order, and every comparison is a
test for an exact zero.  Table~\ref{tab:symbolic} lists the checks; all
passed.  The scripts are part of the deposited data.

\begin{table*}
\footnotesize
\caption{\label{tab:symbolic}Computer-algebra checks.  $N$: number of subsystem
states; $N_b$: number of bath coordinates; ``real'' and ``complex'': real
symmetric and complex Hermitian $h$; ``both conventions'': traceless and
untraced.}
\begin{ruledtabular}
\begin{tabular}{>{\raggedright\arraybackslash}p{0.52\textwidth}>{\raggedright\arraybackslash}p{0.29\textwidth}>{\raggedright\arraybackslash}p{0.12\textwidth}}
Statement & Cases & Outcome\\
\colrule
Eq.~(\Mref{eq:mqcle-full}) on $\rho^{\rm phys}$ reproduces Eq.~(\Mref{eq:qcle}) in every matrix element &
$N=2,3$, $N_b=1$; $N=2$, $N_b=2$; real and complex; both conventions & exact zero\\
Real-symmetric form of $i\cL_m'$ with complex $h$ & $N=2$ & fails, as it must\\
Mixed convention ($h$ in $H_m$, $\bh$ in $i\cL_m'$) & $N=2$ & closure lost\\
Normalization, Eq.~\eqref{eq:wick-norm}, and $H_m-P^2/(2M)-V_b=\sum h_{\lambda\lambda'}(\hat a^\dagger_\lambda\hat a_{\lambda'})_W$ &
$N=2$; real and complex & exact\\
Hermite orders of the image of $\rho^{\rm phys}$: $i\cL_m$ $\{0,2\}$; $i\cL_m^{\PB}$ $\{0,2,4\}$; $i\cL_m'$ alone $\{2,4\}$ although its polynomial degrees are $\{0,2,4\}$ &
$N=2,3$, $N_b=1,2$; real and complex & as stated\\
Repeated application: highest degree $2,2,2$ (exact) and $4,6,8$ (Poisson bracket) &
$N=2$, $1$ to $\symPowers$ applications & as stated\\
Every $\cV_K$ invariant under $i\cL_m$; $i\cL_m^{\PB}$ raises each basis function by exactly two &
all $\symVkCount$ Hermite functions of orders $0$ to $\symVkMax$, $N=2$ & as stated\\
Image of $\rho^{\rm phys}$ in the physical sector: $i\cL_m$ yes, $i\cL_m^{\PB}$ no &
$N=2,3$, $N_b=1,2$; real and complex & as stated\\
Eqs.~(\Mref{eq:raise}), (\Mref{eq:force-split}) and (\Mref{eq:T}) &
all $|\n|\le\symIdOrder$, $N=2$: $\symIdCases$ cases of $(\n,i,j)$ & exact\\
$-i\cL_m$ on $c(R,P)\varphi\He_{\n}$ equals Eqs.~(\Mref{eq:coeff})--(\Mref{eq:RG}); $-i\cL_m^{\PB}$ alone equals Eq.~(\Mref{eq:PBcoeff}) &
all $\symOpBasis$ functions with $|\n|\le\symOpOrder$, $N=2$; real and complex & exact\\
PBME deviation on $\rho^{\rm phys}$: orders $\{2,4\}$; order two is Eq.~(\Mref{eq:inframe}) (traceless) or Eq.~\eqref{eq:raised-trace} (untraced); order four is $\wick{q^{(4)}}$, occupying $\nQuarticModes$ of $\symOrderFourTotal$ functions; $q^{(4)}\ne\wick{q^{(4)}}$ &
$N=2$; real and complex; both conventions & as stated\\
\end{tabular}
\end{ruledtabular}
\par\smallskip{\footnotesize Alt text: a table of eleven symbolic checks with the cases covered and the outcome
of each.}
\end{table*}

The sparse matrices of the production code were also checked against
\Meq{eq:force-split} in the basis of all even orders up to four, in which the
coupling pattern of Fig.~\Mref{fig:ladder}(b) is drawn: the product with
$z_iz_j-\delta_{ij}$ equals $\mathsf T^{ij}$ plus the raising matrix
$\mathsf U^{ij}$ element by element.

\section{Numerical methods and convergence}
\label{sec:S6}

\subsection{Common setting}

Atomic units are used throughout.  The initial partial Wigner function is
\begin{equation}
\rho_W^{\lambda\lambda'}(R,P,0)=\delta_{\lambda1}\delta_{\lambda'1}\,
\frac{1}{\pi\hbar}\exp\Bigl[-\frac{(R-R_0)^2}{2\sigma_R^2}
-\frac{(P-P_0)^2}{2\sigma_P^2}\Bigr],\qquad \sigma_P=\frac{\hbar}{2\sigma_R},
\label{eq:rho0}
\end{equation}
the Wigner function of the minimum-uncertainty packet of the main text.  All
phase-space calculations use uniform periodic grids in $R$ and $P$,
derivatives by fast Fourier transform with the Nyquist component removed, and
the rectangle rule for integrals.  The grids of the coefficient propagation
and of the Strang solver place nodes at $R_{\min}+i\Delta R$, those of the
four-field solver at cell centres.  Where fields are compared point by point
(Sec.~\ref{sec:S7-phase}), the four-field solution is first moved onto the
nodes of the coefficient propagation by an exact Fourier translation.

\subsection{Coefficient propagation}

The semi-discrete operator $\mathbb L$ on the right-hand side of \Meq{eq:coeff}
is applied without
forming a matrix: $\mathsf R(R)$ and $\mathsf G(R)$ are $11\times11$ matrices
built once at each grid point $R$, and the derivatives are applied by Fourier
transform.  A substep of length $\tau$ replaces the state $v$ by
$\beta V_me^{\tau H_m}\bm e_1$, where $\beta=\|v\|$, the columns of $V_m$ span
the Krylov space $\{v,\mathbb Lv,\dots,\mathbb L^{m-1}v\}$ and $H_m$ is the
Arnoldi Hessenberg matrix.\cite{saad1992,hochbruck1997}  In our implementation,
the Arnoldi process uses classical Gram--Schmidt orthogonalization and a second
pass when the first loses more than a factor $\sqrt2$ in norm.  The
substep is accepted when the endpoint Arnoldi residual norm,\cite{botchev2013}
$\beta\,h_{m+1,m}\,|(e^{\tau H_m})_{m1}|$, is below $\krylovTol\,\beta$;
otherwise $\tau$ is reduced by the factor
$\max\{0.2,\;0.9\,(\krylovTol\,\beta/{\rm err})^{1/m}\}$ and the estimate
repeated with the same basis.  We use the residual to choose the substep and
check the propagation error independently with the Runge--Kutta calculation
below.  After an accepted substep $\tau$ may grow by at
most a factor of five.  A breakdown of the Arnoldi process makes the substep
exact.  The subspace dimension, $m=12$ ($8$ at $P_0=10$), is limited by the
memory needed for $m+1$ copies of the state, not by accuracy.

As an independent check of the propagator, the same semi-discrete equations
were integrated with the classical fourth-order Runge--Kutta method at a
sequence of step sizes, at $P_0=40$ up to $t=\ghRkT$ on a $\ghRkGrid$ grid.
The Runge--Kutta solution converges to the exponential propagation at order
$\ghRkOrder$, and at the smallest step, $\ghRkDt$, the two differ by
$\ghRkErr$ relative to the norm of the full coefficient state.

\subsection{Four-field solver of the subsystem QCLE}

With $A=\rho_W^{11}$, $C=\rho_W^{22}$, $\rho_W^{12}=b_R+ib_I$, $V=V_b\one+h$
and $V_0'=\tfrac12(\partial_RV_{11}+\partial_RV_{22})$, \Meq{eq:qcle} for real
symmetric $h$ reads
\begin{equation}
\begin{aligned}
\partial_tA&=-\frac PM\partial_RA+\partial_RV_{11}\,\partial_PA
+\partial_RV_{12}\,\partial_Pb_R-\frac{2h_{12}}{\hbar}\,b_I,\\
\partial_tC&=-\frac PM\partial_RC+\partial_RV_{22}\,\partial_PC
+\partial_RV_{12}\,\partial_Pb_R+\frac{2h_{12}}{\hbar}\,b_I,\\
\partial_tb_R&=-\frac PM\partial_Rb_R+V_0'\,\partial_Pb_R
+\frac12\partial_RV_{12}\,\partial_P(A+C)+\frac{h_{11}-h_{22}}{\hbar}\,b_I,\\
\partial_tb_I&=-\frac PM\partial_Rb_I+V_0'\,\partial_Pb_I
+\frac{h_{12}(A-C)-(h_{11}-h_{22})\,b_R}{\hbar}.
\end{aligned}
\label{eq:fourfield}
\end{equation}
These four real fields are advanced with the classical fourth-order
Runge--Kutta method.  It is a separate code that uses neither the right-hand side nor the
propagator of the other two solvers; on a common grid its right-hand side
agrees with the subsystem-QCLE operator of the Strang code to round-off.  The step is below the stability limit set by the largest
advection and force frequencies of the grid: $\gridDtSmall$ at
$P_0=\gridDtSmallList$ and $\gridDtLarge$ otherwise.  It uses the grids of
Table~\Mref{tab:params}, except at $P_0=10$, where it was run on a grid
coarser by two in each direction.  Table~\ref{tab:convergence} lists the
changes of $\rho^{22}(T)$ under halving of its step and of its grid
resolution.

\subsection{Strang splitting}

The Strang solver splits \Meq{eq:qcle} into the bath advection
$-(P/M)\partial_R$, diagonal in the Fourier variable $k_R$ at fixed $P$, and
the combined electronic and force terms, diagonal in $k_P$ at fixed $R$.  In
the second, with $\partial_P\to ik_P$,
\begin{equation}
\partial_t\hat\rho=-i\mathsf A\hat\rho+i\hat\rho\,\mathsf B,\qquad
\mathsf A=\frac{h}{\hbar}+\frac{k_P}{2}F,\quad
\mathsf B=\frac{h}{\hbar}-\frac{k_P}{2}F,
\label{eq:strang}
\end{equation}
with $F=-\partial_R(V_b\one+h)$, and the exact solution is
$\hat\rho(t)=e^{-i\mathsf At}\hat\rho(0)e^{i\mathsf Bt}$ with $\mathsf A$ and
$\mathsf B$ real symmetric.  Both sub-flows are therefore integrated exactly;
every factor is unitary, so the scheme is stable at any step, conserves the
trace ($\mathsf A=\mathsf B$ at $k_P=0$) and preserves Hermiticity.  The only
error is that of the splitting, of second order in the step.  At the step
$0.5$ its maximum deviation over time from the four-field solution falls by
factors between $\strangHalveMin$ and $\strangHalveMax$ when the step is
halved [Fig.~\ref{fig:convergence}(a)].  At $P_0=10$ the two were compared on
the coarser grid of the four-field solver, where they differ by
$\strangLinfPTen$ whether the Strang step is $0.5$ or $0.25$; this difference
is set by the grid and does not resolve the splitting error.

\subsection{Full quantum reference}

The two-component wave function is propagated by the split-operator
method,\cite{feit1982} using a Fourier grid representation.\cite{kosloff1983}
Half steps of the kinetic energy in
momentum space surround a full step of the electronic propagator, obtained at
each grid point from the eigenvectors of $h(R)$.  Production runs use
$\qNRProdLow$ grid points ($\qNRProdHigh$ for $P_0\ge70$) and the step
$\qDtProd$.  Halving the step twice gives the observed order of the splitting,
between $\qOrderMin$ and $\qOrderMax$, and a Richardson estimate of the error
remaining at the finest step; separate runs double the grid density and
enlarge the box.  The value used as the reference comes from a run that
combines all three refinements, with $\qNRRefLow$ grid points
($\qNRRefHigh$ for $P_0\ge70$) and the step $\qDtRef$; its uncertainty
is the largest of the Richardson estimate and the two spatial changes
(Table~\ref{tab:convergence}).  The
norm is conserved to $\qNormDrift$.  Adiabatic populations are obtained by
projecting the wave function on the eigenvectors of $h(R)$ at each grid point.

\subsection{PBME estimators}

\emph{Projected density.}  For the initial state $\lambda=1$ the physical
density is $\phi\,\Pi_{11}\,\rho_W^{11}(R,P,0)$, with
$\Pi_{11}=2|w_1|^2-1$ by Eq.~\eqref{eq:Pi-w}.  Under $\varphi$, $u=|w_1|^2$ is
exponentially distributed with unit mean, so
$Z=\langle|\Pi_{11}|\rangle=\int_0^\infty du\,e^{-u}|2u-1|=4e^{-1/2}-1$.  Mapping
variables are drawn from $\varphi|\Pi_{11}|/Z$ and bath variables from
$\rho_W^{11}(R,P,0)$, and sample $s$ carries the weight
$Z\,{\rm sgn}\,\Pi_{11}(x_s)$; a fraction
$(2e^{-1/2}-1)/Z=0.149$ of the samples carries a negative weight, as found in
the calculations ($\pbNegFrac$).  Each sample is propagated with Hamilton's
equations for $H_m$ by a symmetric step: half a step of the bath momentum
under $-\partial_RH_m$, half a step of the bath coordinate, the exact rotation
of the mapping variables at fixed $R$, and the same two half steps in reverse
order.  The step is reversible to round-off.

The PBME value of $\int dR\,dP\,\rho_W^{k'k}(t)$ is the phase-space average of
$(\hat a^\dagger_k\hat a_{k'})_W$ over the PBME density, which at $t=0$ equals
it by Eq.~\eqref{eq:wick-norm}.  Because Hamiltonian flow conserves
phase-space volume, it is estimated without bias by
\begin{equation}
\hat\rho^{k'k}(t)=\frac{Z}{N_s}\sum_{s=1}^{N_s}{\rm sgn}\,\Pi_{11}\bigl(x_s(0)\bigr)\,
\bigl(\hat a^\dagger_k\hat a_{k'}\bigr)_W\bigl(x_s(t)\bigr).
\label{eq:estimator}
\end{equation}
The weights are not renormalized by their sample sum, which would introduce a
finite-sample bias.  The control variate
$\hat\rho(t)-\hat\rho(0)+\rho(0)$, with $\rho(0)$ the exact initial density
matrix, is also unbiased and has a smaller variance.

\emph{Statistics.}  The samples are divided into blocks of $\pbBlockSize$;
the quoted standard error is that of the mean over the $\pbNBlocks$ blocks
($\pbNBlocksMin$ at $P_0=20$).  The blocks are generated in \numword{\pbNSeeds}
(\numword{\pbNSeedsMin}) independently seeded groups, and the standard error computed
from the scatter between the groups is between $\semRatioMin$ and
$\semRatioMax$ times the block value, as expected of an estimate with so few
groups.  Computed from nested subsets of four or more blocks, the block
standard error decreases with the number of samples $N$ with log--log slopes
between $\pbSlopeMin$ and $\pbSlopeMax$ [Fig.~\ref{fig:convergence}(c)].  At
$P_0=40$, with $\pbDtSamples$ trajectories run from the same initial
ensemble at the steps $0.5$, $0.25$ and $0.125$, the observed order of the
step is $\pbDtOrder$ and halving the production step changes $\rho^{22}(T)$ by
$\pbDtChange$, a fraction $\pbDtOverSem$ of the standard error of the
production estimate.

\emph{Focused sampling.}  Each mapping oscillator starts on the circle
$r_\lambda^2+p_\lambda^2=\hbar(2\delta_{\lambda1}+1)$ with an independent
uniform phase, the bath variables are drawn from Eq.~\eqref{eq:rho0}, and the
density matrix is read out with the symbols $(\hat a^\dagger_k\hat a_{k'})_W$;
the population symbols equal $\delta_{k1}$ on the initial
circles.\cite{bonella2003}  The same trajectory integrator is used.  The
adiabatic populations and the coherence of this ensemble were obtained by
propagating it a second time from the same seed, which reproduces its
diabatic populations to within $\focReproMax$.

\subsection{Summary of numerical uncertainties}
\label{sec:S6-summary}

\begin{figure*}
\centering
\includegraphics[width=\textwidth]{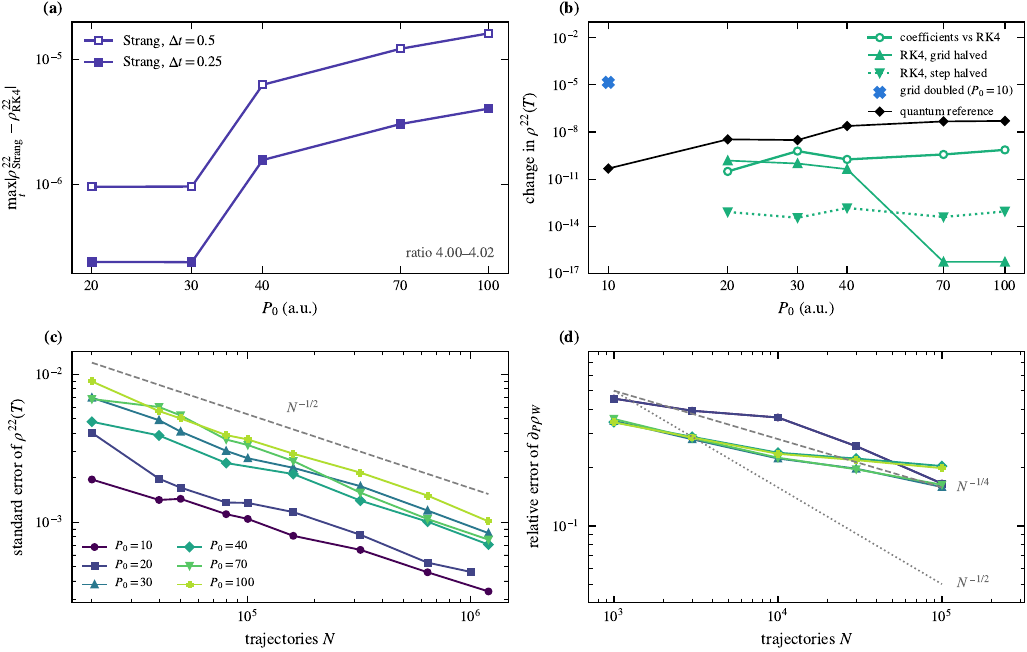}
\caption{\label{fig:convergence}Convergence of the reference calculations and of the
estimators.  (a) Largest difference over time between the Strang and RK4
solutions at two steps.  (b) Changes of $\rho^{22}(T)$ in the tests of
Table~\ref{tab:convergence}.  (c) Standard error of the projected-density
PBME estimate of $\rho^{22}(T)$ against the number of trajectories.  (d)
Relative error of a kernel estimate of $\partial_P\rho_W$ from $N$ samples;
lines: $N^{-1/4}$ and $N^{-1/2}$.}
\par\smallskip{\footnotesize Alt text: four log--log panels showing a factor-four drop of the Strang error on
halving the step, refinement changes far below the physical errors,
statistical errors falling as the inverse square root of the sample size, and
a slowly decreasing error of the kernel gradient estimate.}
\end{figure*}

Table~\ref{tab:convergence} collects the refinement tests, and
Fig.~\ref{fig:convergence}(b) shows them graphically.  The first two columns
give the difference in $\rho^{22}$ between the coefficient propagation and the
four-field (RK4) solution on a common grid, at $t=T$ and as the largest
difference over the instants recorded by both.  The next two give the change
of the RK4 value at $T$ when its step is halved and when its grid is coarsened
by two in each direction, and the two after them the largest difference over
time between the Strang and RK4 solutions at the steps $0.5$ and $0.25$.  The
last two are the adopted uncertainties, $u_{\rm QCLE}$ of the QCLE value and
$u_{\rm quantum}$ of the quantum reference.  Changes near $10^{-16}$ are at the
level of round-off.

For $P_0\ge20$ we take as $u_{\rm QCLE}$ the largest of three numbers: the
difference at $T$ between the coefficient propagation and the four-field
solver on their common grid, and the changes of the four-field value under
halving of its step, at most $\gridDtHalfMax$, and of its grid resolution, at
most $\gridSpatialMax$.  Because the grid is common to both solvers, the last
test also resolves the grid of the coefficient propagation.  The resulting
$u_{\rm QCLE}$ is at most $\numUncHigh$, and the quantum reference is uncertain
by at most $\qUncMax$; both lie far below the physical differences
$E_{\rm QCLE}$ and $E_{\rm PBME}$.

At $P_0=10$ the slow packet spends a long time in the interaction region, the
state-$2$ component develops fine structure in momentum, and the calculation
is correspondingly demanding.  On the $\ghGridPTen$ grid of the coefficient
propagation, a fraction of at most $\pTenBoundary$ of the density comes within
one cell of the edge of the box, against $\pTenBoundaryCoarse$ on a grid
coarser by two in each direction; the change of $\rho^{22}$ between the two
grids, computed with the Strang solver, is at most $\numUncPTen$ over the
whole propagation.  We take this as $u_{\rm QCLE}$ at $P_0=10$; $E_{\rm QCLE}$
exceeds it by a factor $\qcleOverNumPTen$.
The four-field solver was run at $P_0=10$ only on the coarser grid, where it
differs from the Strang solution by $\strangLinfPTen$, and on a grid coarser
still by two its result changes by $\gridCoarsePTen$.  Its differences from
the coefficient propagation in Table~\ref{tab:convergence} compare solutions
on different grids; they are not a test of the solvers and are not shown in
Fig.~\Mref{fig:numerics}(b).

\begin{table*}
\footnotesize
\caption{\label{tab:convergence}Numerical accuracy of the final population
$\rho^{22}(T)$: differences between the solvers of the subsystem QCLE, changes
under refinement, and the adopted uncertainties (Sec.~\ref{sec:S6-summary}).
Dashes: test not performed.}
\begin{ruledtabular}
\begin{tabular}{rllllllll}
 & \multicolumn{2}{c}{coefficients $-$ RK4} & \multicolumn{2}{c}{RK4 refinement}
 & \multicolumn{2}{c}{Strang $-$ RK4} & & \\
\cline{2-3}\cline{4-5}\cline{6-7}
$P_0$ & $t=T$ & all $t$ & $\Delta t/2$ & grid$/2$ & $\Delta t=0.5$ &
$\Delta t=0.25$ & $u_{\rm QCLE}$ & $u_{\rm quantum}$\\
\colrule
10  & ${\eGridPTen}^{a}$ & ${\eGridMaxPTen}^{a}$ & -- & ${\gridCoarsePTen}^{a}$ & ${\strangLinfPTen}^{a}$ & ${\strangLinfHalfPTen}^{a}$ & ${\numUncPTen}^{b}$ & $\qUncPTen$\\
20  & $\eGridPTwenty$ & $\eGridMaxPTwenty$ & $\gridDtHalfPTwenty$ & $\gridSpatialPTwenty$ & $\strangLinfPTwenty$ & $\strangLinfHalfPTwenty$ & $\numUncPTwenty$ & $\qUncPTwenty$\\
30  & $\eGridPThirty$ & $\eGridMaxPThirty$ & $\gridDtHalfPThirty$ & $\gridSpatialPThirty$ & $\strangLinfPThirty$ & $\strangLinfHalfPThirty$ & $\numUncPThirty$ & $\qUncPThirty$\\
40  & $\eGridPForty$ & $\eGridMaxPForty$ & $\gridDtHalfPForty$ & $\gridSpatialPForty$ & $\strangLinfPForty$ & $\strangLinfHalfPForty$ & $\numUncPForty$ & $\qUncPForty$\\
70  & $\eGridPSeventy$ & $\eGridMaxPSeventy$ & $\gridDtHalfPSeventy$ & $\gridSpatialPSeventy$ & $\strangLinfPSeventy$ & $\strangLinfHalfPSeventy$ & $\numUncPSeventy$ & $\qUncPSeventy$\\
100 & $\eGridPOneHundred$ & $\eGridMaxPOneHundred$ & $\gridDtHalfPOneHundred$ & $\gridSpatialPOneHundred$ & $\strangLinfPOneHundred$ & $\strangLinfHalfPOneHundred$ & $\numUncPOneHundred$ & $\qUncPOneHundred$\\
\end{tabular}
\end{ruledtabular}
\raggedright
$^{a}$Four-field solver on a grid coarser by two in each direction than that
of Table~\Mref{tab:params}.  The coefficients $-$ RK4 entries compare
solutions on different grids and are not shown in Fig.~\Mref{fig:numerics}(b);
the Strang $-$ RK4 entries are computed on the coarser grid; the grid$/2$ test
coarsens it by two again.\\
$^{b}$Change of the Strang solution between the grid of
Table~\Mref{tab:params} and one coarser by two in each direction, largest over
time.
\par\smallskip{\footnotesize Alt text: a table of solver differences, refinement changes and adopted
uncertainties for six initial momenta.}
\end{table*}

\section{Additional results}
\label{sec:S7}

\subsection{Closure in enlarged bases}

\begin{figure}
\centering
\includegraphics[width=0.55\textwidth]{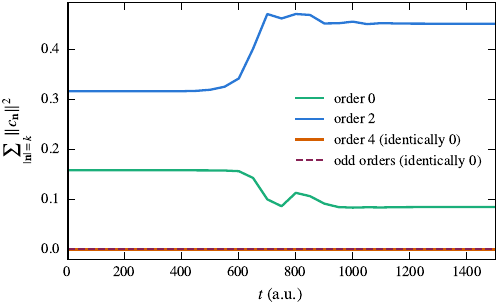}
\caption{\label{fig:closure}Content of each Hermite order in the propagation of
physical initial data at $P_0=\clPZero$: even orders up to four, and odd
orders in the basis of all orders up to three.}
\par\smallskip{\footnotesize Alt text: a line plot against time in which the order-zero and order-two contents
are positive and the order-four and odd contents are zero throughout.}
\end{figure}

The closure test of Sec.~\Mref{sec:closure-results} propagates the
coefficient equations at $P_0=\clPZero$ on a $\clGrid$ grid in the basis of
all even orders up to four ($\clModes$ modes) from the physical initial data
of Eq.~\eqref{eq:rho0}.  Figure~\ref{fig:closure} shows the content of each
order, $E_k=\sum_{|\n|=k}\int dR\,dP\,|c_{\n}|^2$.  The order-four content is
zero to the last bit at every recorded time while the order-zero content
changes as the packet splits at the crossing.  In the control run every
order-four coefficient starts as the Gaussian packet of
Eq.~\eqref{eq:rho0} centred in the interaction region at $R=0$,
multiplied by an independent random amplitude;
this gives $E_4=\clCtlEFour$ and $E_2=0$ at $t=0$, and $E_2=\clCtlETwo$ at
$t=\clCtlT$.  The parity test uses the basis of all orders up to three
($\parModes$ modes, $\parModesOdd$ of odd order) in the same way: the odd
content stays exactly zero for physical data, and in the control run odd data
with $E_{\rm odd}=\parCtlEOdd$ evolve to $E_{\rm odd}=\parCtlEOddFinal$ at
$t=\clCtlT$ without any even content having appeared.

\subsection{Phase-space densities}
\label{sec:S7-phase}

\emph{Fields.}  The quantum field is the partial Wigner transform of the
converged wave function at $t=T$,
\begin{equation}
\rho_W^{\lambda\lambda'}(R,P)=\frac{1}{2\pi\hbar}\int dy\,e^{-iPy/\hbar}\,
\psi_\lambda(R+y/2)\,\psi^*_{\lambda'}(R-y/2),
\label{eq:partial-wigner}
\end{equation}
evaluated on the nodes of the coefficient propagation; its trace differs from
one by at most $\psTraceErr$.  The QCLE field is read out of the coefficient
propagation on the same nodes.  Two independent solutions of the subsystem
QCLE are compared with it: the Strang solution at the step $\psStrangStep$, on
the same nodes, and the four-field Runge--Kutta solution at its production
step, whose nodes lie at cell centres and which is moved onto the nodes of
the coefficient propagation by an exact Fourier translation by half a cell in
each direction.  The square root of the fraction of spectral power beyond
three quarters of the Nyquist wavenumber is at most $\psTailMax$ for the
quantum and QCLE fields, so the grids resolve them.  At $P_0=10$ they do not:
for the QCLE densities at $t=T$ this quantity is at least $\psTailPTen$,
although the populations are converged (Sec.~\ref{sec:S6}), and that regime
is not compared.

\emph{Solutions of the subsystem QCLE.}  The Strang solution differs from the
coefficient propagation by at most $\psStrangPeakMax$ of the peak density in
all four regimes, the four-field solution by at most $\psGridPeakPOneHundred$,
at $P_0=100$; the values for each regime are collected at the end of this
section.  The difference of the four-field solution can be predicted.  Where the forces and the
coupling vanish, each Fourier component $\hat\rho_W^{\lambda\lambda'}(k,P)$ of
the density in $R$ is multiplied over a step $\Delta t$ by $e^{\zeta}$, with
$\zeta=-ikP\Delta t/M$ for the diagonal densities and
$\zeta=(-ikP/M+iE_0/\hbar)\Delta t$ for $\rho_W^{12}$, $E_0$ being the asymptotic
gap $V_{22}-V_{11}$; the Runge--Kutta scheme replaces $e^{\zeta}$ by
$G(\zeta)=\sum_{n=0}^{4}\zeta^n/n!$.  After $N=T/\Delta t$ steps the four-field
density therefore differs from the exact one by the field with Fourier
components $\{[G(\zeta)e^{-\zeta}]^{N}-1\}\,\hat\rho_W^{\lambda\lambda'}$.  Evaluated
with the coefficient-propagation density, this prediction neglects the forces
and the coupling near the crossing.  At $P_0=\psRkList$ it reproduces the
difference of every component to within a relative $L_2$ residual of
$\psRkResidMax$, and the remainder is at most $\psRkAfterMax$ of the peak.  At
$P_0=20$ the predicted error, $\psRkPredPTwenty$ of the peak, is smaller than
the measured difference, $\psGridPeakPTwenty$, which is of the size of the
difference between the Strang solution and the coefficient propagation,
$\psStrangPeakPTwenty$.  Since $G(0)=1$, the free-flight error leaves the
integral over $R$ of each diagonal density unchanged, and with it the
populations and the momentum distributions.

\emph{PBME cells.}  The estimator of \eqref{eq:estimator} is resolved in phase
space by summing the weighted readouts of the trajectories found in each cell
at $t=T$ and dividing by the cell area.  The trajectories are those of the
population estimates of the main text: propagated again with the same seeds,
sampler and integrator, they reproduce the archived totals of each group of
samples to $\psReproMax$.  The cells are blocks of grid cells with edges on
grid nodes, of $\psCellRPTwenty\times\psCellPPTwenty$, $\psCellRPThirty\times
\psCellPPThirty$, $\psCellRPSeventy\times\psCellPPSeventy$ and
$\psCellRPOneHundred\times\psCellPPOneHundred$ in $R$ and $P$ at $P_0=20$, $30$,
$70$ and $100$; at $P_0=20$ and $30$ they are enlarged to hold the parts of
the QCLE and PBME densities that lie away from the packets.  They
contain all but a fraction $\psTrajOutsideMax$ of the trajectories, which
could change a distance $D_a$ by at most $\psOutsidePbmeMax$, and all but a
fraction $\psOutsideMax$ of $|\rho_W|$ of every grid field.  The standard
error of a cell follows from independent blocks of trajectories.  A grid
field is reduced to a cell by its exact average over the cell, obtained by
applying the Fourier multiplier $e^{ikh/2}\sin(kh/2)/(kh/2)$, with $h$ the
cell width, along each axis; this reproduces the cell averages of a resolved
Gaussian to round-off.

\emph{Measures.}  The negative volume $N_a$ defined in the main text is
evaluated on the grid for the quantum and QCLE densities; the momentum
distance $D_a$ is evaluated on the cells for all three.  We also give the
relative $L_2$ distance $\|X-Y\|/\|Y\|$ between cell averages.  For
the PBME, statistical noise adds to every distance; its contribution is
estimated from the blocks as $(2/\pi)^{1/2}\sum_l s_l\,\Delta P_c/\rho^{aa}$
for $D_a$, with $s_l$ the standard error of the momentum distribution in
cell column $l$ of width $\Delta P_c$, and as $(\sum s^2)^{1/2}/\|Y\|$ for the
$L_2$ distances.  Every PBME value of $D_a$ exceeds this floor by a factor of
at least $\psMomOverFloorMin$, and every PBME $L_2$ distance by a factor of at
least $\psLtwoOverFloorMin$.

\begin{figure*}
\centering
\includegraphics[width=\textwidth]{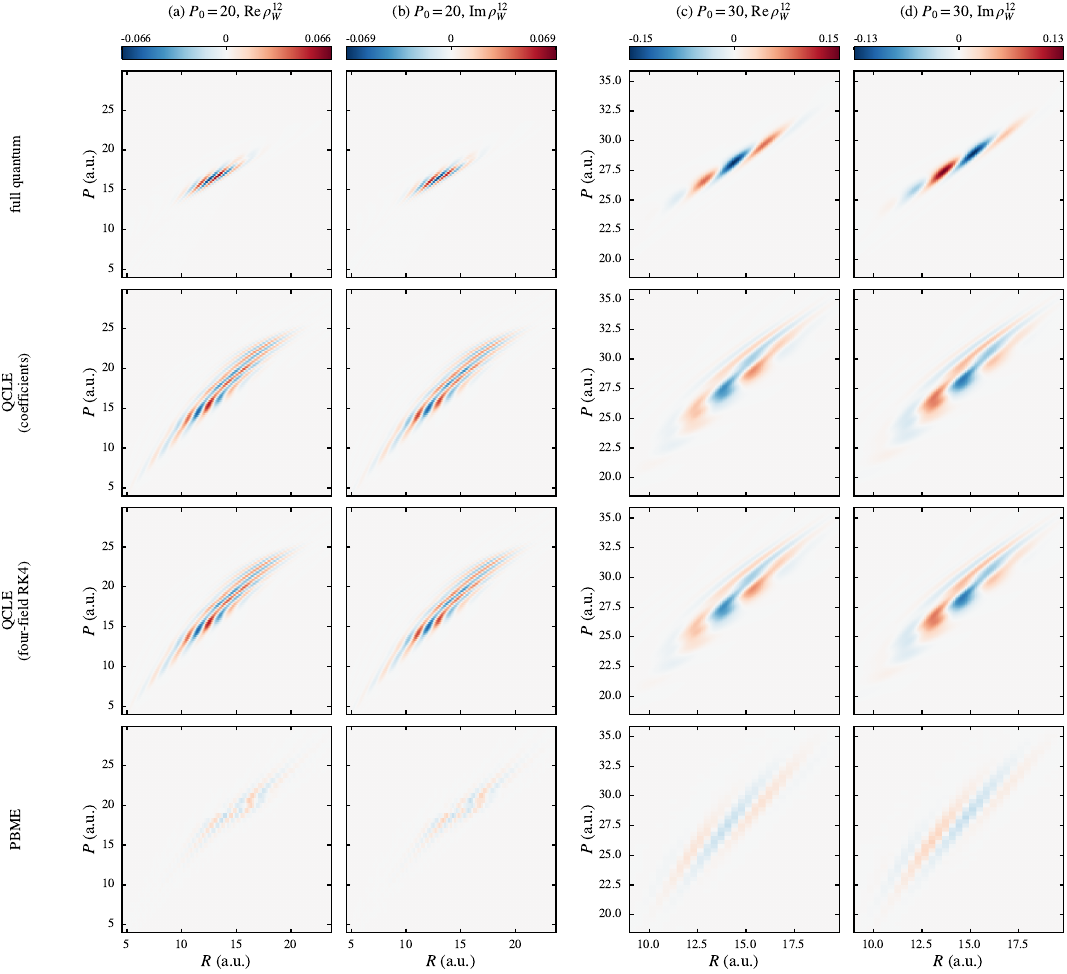}
\caption{\label{fig:phase-coherence}Real and imaginary parts of $\rho_W^{12}$ at
$t=T$ for $P_0=20$ and $30$, arranged as in Fig.~\Mref{fig:phasespace}.}
\par\smallskip{\footnotesize Alt text: sixteen colour maps of the coherence in the position--momentum plane
from four calculations; the exact coherence is a compact fringe pattern
between the two packets, the QCLE coherence a longer band, and the PBME one a
noisier pattern.}
\end{figure*}

\begin{figure*}
\centering
\includegraphics[width=\textwidth]{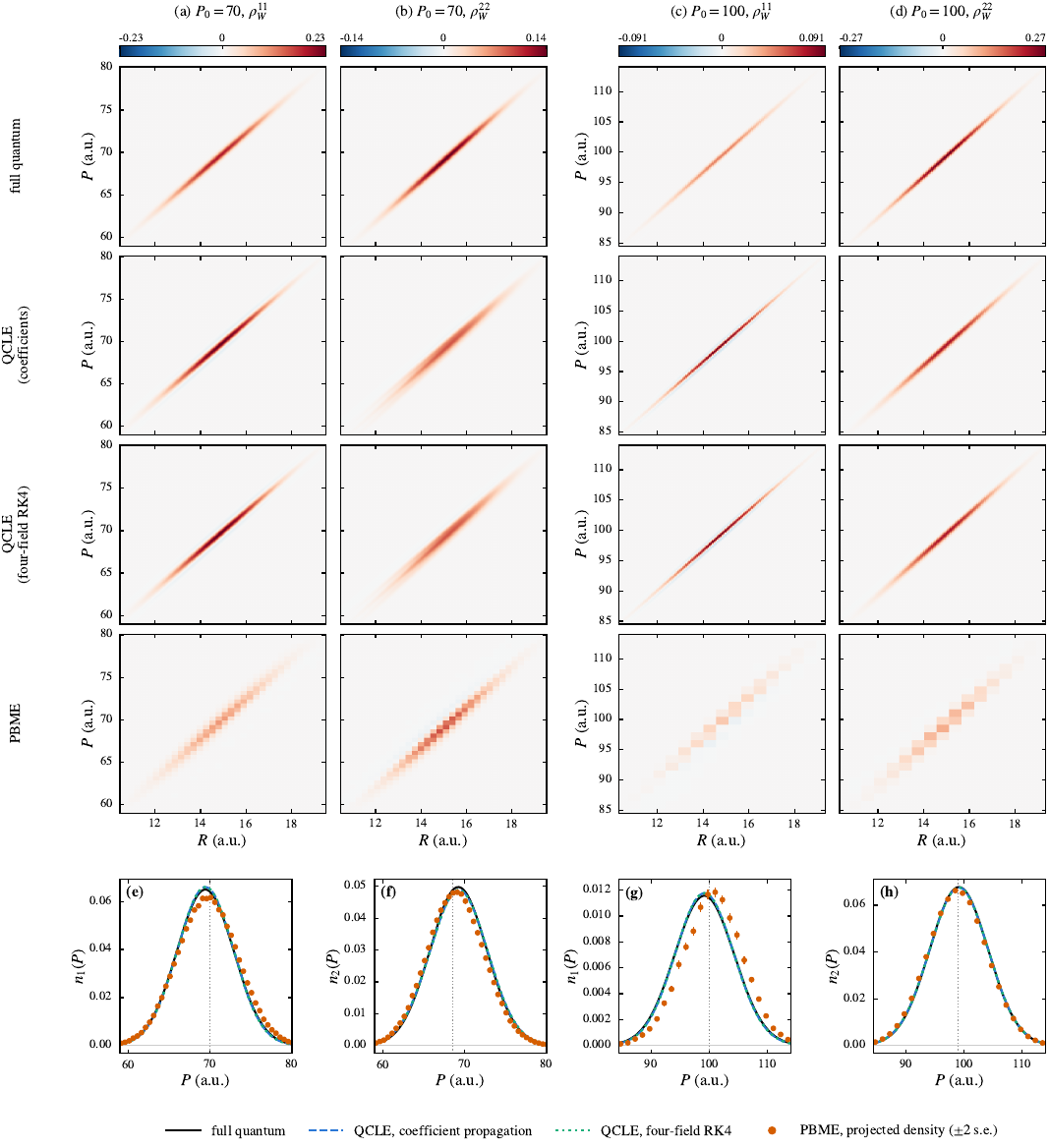}
\caption{\label{fig:phase-high}Diagonal densities and momentum distributions at
$t=T$ for $P_0=70$ and $100$, arranged as in Fig.~\Mref{fig:phasespace}.}
\par\smallskip{\footnotesize Alt text: sixteen colour maps of the two diagonal densities from four calculations
and four panels of momentum distributions; the QCLE follows the exact densities
closely, and the PBME momentum distribution of state one is displaced to
higher momentum.}
\end{figure*}

Tables~\ref{tab:phase-neg}, \ref{tab:phase-mom} and \ref{tab:phase-l2} collect the measures for the
four regimes, Table~\ref{tab:phase-solvers} the differences between the three
solutions of the subsystem QCLE, and Figs.~\ref{fig:phase-coherence} and
\ref{fig:phase-high} show the coherence at $P_0=20$ and $30$ and the diagonal
densities at $P_0=70$ and $100$.

\begin{table}
\caption{\label{tab:phase-neg}Negative volumes $N_a$ of the diagonal densities at
$t=T$, and the asymptotic momentum difference $\Delta P$ of the two surfaces in
units of $\sigma_P$.}
\begin{ruledtabular}
\begin{tabular}{rrllll}
 & & \multicolumn{2}{c}{$N_1$} & \multicolumn{2}{c}{$N_2$}\\
\cline{3-4}\cline{5-6}
$P_0$ & $\Delta P/\sigma_P$ & quantum & QCLE & quantum & QCLE\\
\colrule
20  & $\psShiftPTwenty$ & $\psNegQOnePTwenty$ & $\psNegQcleOnePTwenty$ & $\psNegQTwoPTwenty$ & $\psNegQcleTwoPTwenty$\\
30  & $\psShiftPThirty$ & $\psNegQOnePThirty$ & $\psNegQcleOnePThirty$ & $\psNegQTwoPThirty$ & $\psNegQcleTwoPThirty$\\
70  & $\psShiftPSeventy$ & $\psNegQOnePSeventy$ & $\psNegQcleOnePSeventy$ & $\psNegQTwoPSeventy$ & $\psNegQcleTwoPSeventy$\\
100 & $\psShiftPOneHundred$ & $\psNegQOnePOneHundred$ & $\psNegQcleOnePOneHundred$ & $\psNegQTwoPOneHundred$ & $\psNegQcleTwoPOneHundred$\\
\end{tabular}
\end{ruledtabular}
\par\smallskip{\footnotesize Alt text: a table of negative volumes of the quantum and QCLE densities of both
states for four initial momenta.}
\end{table}

\begin{table}
\caption{\label{tab:phase-mom}Distance $D_a$ of the momentum distribution of state
$a$ from the exact one, in units of the exact population.}
\begin{ruledtabular}
\begin{tabular}{rllll}
 & \multicolumn{2}{c}{$D_1$} & \multicolumn{2}{c}{$D_2$}\\
\cline{2-3}\cline{4-5}
$P_0$ & QCLE & PBME & QCLE & PBME\\
\colrule
20  & $\psMomQcleOnePTwenty$ & $\psMomPbmeOnePTwenty$ & $\psMomQcleTwoPTwenty$ & $\psMomPbmeTwoPTwenty$\\
30  & $\psMomQcleOnePThirty$ & $\psMomPbmeOnePThirty$ & $\psMomQcleTwoPThirty$ & $\psMomPbmeTwoPThirty$\\
70  & $\psMomQcleOnePSeventy$ & $\psMomPbmeOnePSeventy$ & $\psMomQcleTwoPSeventy$ & $\psMomPbmeTwoPSeventy$\\
100 & $\psMomQcleOnePOneHundred$ & $\psMomPbmeOnePOneHundred$ & $\psMomQcleTwoPOneHundred$ & $\psMomPbmeTwoPOneHundred$\\
\end{tabular}
\end{ruledtabular}
\par\smallskip{\footnotesize Alt text: a table of momentum-distribution distances of the QCLE and the PBME for
both states and four initial momenta.}
\end{table}

\begin{table}
\caption{\label{tab:phase-l2}Relative $L_2$ distance $\|X-Y\|/\|Y\|$ of the cell
averages of $\rho_W^{11}$, $\rho_W^{22}$ and ${\rm Re}\,\rho_W^{12}$ from the
exact densities.}
\begin{ruledtabular}
\begin{tabular}{rllllll}
 & \multicolumn{2}{c}{$\rho_W^{11}$} & \multicolumn{2}{c}{$\rho_W^{22}$} &
\multicolumn{2}{c}{${\rm Re}\,\rho_W^{12}$}\\
\cline{2-3}\cline{4-5}\cline{6-7}
$P_0$ & QCLE & PBME & QCLE & PBME & QCLE & PBME\\
\colrule
20  & $\psLtwoQcleOnePTwenty$ & $\psLtwoPbmeOnePTwenty$ & $\psLtwoQcleTwoPTwenty$ & $\psLtwoPbmeTwoPTwenty$ & $\psLtwoQcleRePTwenty$ & $\psLtwoPbmeRePTwenty$\\
30  & $\psLtwoQcleOnePThirty$ & $\psLtwoPbmeOnePThirty$ & $\psLtwoQcleTwoPThirty$ & $\psLtwoPbmeTwoPThirty$ & $\psLtwoQcleRePThirty$ & $\psLtwoPbmeRePThirty$\\
70  & $\psLtwoQcleOnePSeventy$ & $\psLtwoPbmeOnePSeventy$ & $\psLtwoQcleTwoPSeventy$ & $\psLtwoPbmeTwoPSeventy$ & $\psLtwoQcleRePSeventy$ & $\psLtwoPbmeRePSeventy$\\
100 & $\psLtwoQcleOnePOneHundred$ & $\psLtwoPbmeOnePOneHundred$ & $\psLtwoQcleTwoPOneHundred$ & $\psLtwoPbmeTwoPOneHundred$ & $\psLtwoQcleRePOneHundred$ & $\psLtwoPbmeRePOneHundred$\\
\end{tabular}
\end{ruledtabular}
\par\smallskip{\footnotesize Alt text: a table of relative distances of the QCLE and PBME densities from the
exact ones for three components and four initial momenta.}
\end{table}

\begin{table}
\caption{\label{tab:phase-solvers}Differences of the RK4 and Strang solutions of the
subsystem QCLE from the coefficient propagation at $t=T$: largest relative
$L_2$ distance over the four components and largest pointwise difference over
the peak density.  The ``after subtraction'' column gives the RK4 difference
after subtracting its predicted free-flight error.}
\begin{ruledtabular}
\begin{tabular}{rlllll}
 & \multicolumn{3}{c}{four-field} & \multicolumn{2}{c}{Strang}\\
\cline{2-4}\cline{5-6}
$P_0$ & $L_2$ & max/peak & \shortstack{after\\subtraction} & $L_2$ & max/peak\\
\colrule
20  & $\psGridLtwoPTwenty$ & $\psGridPeakPTwenty$ & $\psGridCorrPTwenty$ & $\psStrangLtwoPTwenty$ & $\psStrangPeakPTwenty$\\
30  & $\psGridLtwoPThirty$ & $\psGridPeakPThirty$ & $\psGridCorrPThirty$ & $\psStrangLtwoPThirty$ & $\psStrangPeakPThirty$\\
70  & $\psGridLtwoPSeventy$ & $\psGridPeakPSeventy$ & $\psGridCorrPSeventy$ & $\psStrangLtwoPSeventy$ & $\psStrangPeakPSeventy$\\
100 & $\psGridLtwoPOneHundred$ & $\psGridPeakPOneHundred$ & $\psGridCorrPOneHundred$ & $\psStrangLtwoPOneHundred$ & $\psStrangPeakPOneHundred$\\
\end{tabular}
\end{ruledtabular}
\par\smallskip{\footnotesize Alt text: a table of relative distances and pointwise differences of two grid
solutions from the coefficient propagation for four initial momenta.}
\end{table}

\section{Estimating the defect rate from trajectories}
\label{sec:S8}

The rate $\mathcal D_4$ of \Meq{eq:D4} needs the momentum gradient
$\partial_P\rho_W$.  On a grid it is available spectrally; a trajectory
calculation has to estimate it from samples.  We tested this estimate on the
initial distribution in Eq.~\eqref{eq:rho0}, whose momentum gradient is known
analytically.  We used a product Gaussian kernel with bandwidth
$1.06\,\hat\sigma N^{-1/5}$ in each variable, following Silverman's
normal-reference rule with a robust estimate of the spread.\cite{silverman1986}
For each variable, $\hat\sigma=\min\{s,\mathrm{IQR}/1.349\}$, where $s$ is the
sample standard deviation and $\mathrm{IQR}$ is the interquartile range;
$N$ is the number of trajectories.  If the interquartile range is zero, we use
$s$ for the spread.  We set a minimum bandwidth of two grid cells and differentiated
the kernel analytically.  The bandwidths were chosen for the marginal
densities; we tested the resulting gradient estimate against its exact value.
We used focused initial ensembles with uniform weights.  For each ensemble,
we computed the RMS error of $\partial_P\rho_W^{11}$ at grid nodes where
$\rho_W^{11}(R,P,0)$ exceeds $10^{-4}$ of its peak, and divided it by the RMS
of the exact derivative over the same nodes.
Averaged over \numword{\kdeSeeds} independent ensembles, this relative error is
$\kdeRelFirstMin$ to $\kdeRelFirstMax$ at $N=\kdeNMin$ and $\kdeRelLastMin$ to
$\kdeRelLastMax$ at $N=\kdeNMax$, depending on the regime, and decreases as
$N^{\alpha}$ with $\alpha$ between $\kdeExpMin$ and $\kdeExpMax$
[Fig.~\ref{fig:convergence}(d)].  For a sufficiently smooth density in two dimensions and a
second-order kernel such as the Gaussian, with independent samples, the optimal asymptotic mean
integrated squared error of a first-derivative estimate scales as $N^{-1/2}$.
Its square root therefore scales as $N^{-1/4}$,\cite{chacon2011} while the
sampling error of the populations falls as $N^{-1/2}$.  The derivative rate
applies to independent samples from an ordinary density.  For the estimator
and bandwidths used here, we measure convergence directly in Fig.~\ref{fig:convergence}(d).
The slow decrease observed there means that estimating $\mathcal D_4$ from
trajectories requires far larger ensembles than estimating the populations.

\section{Changes of mapping coordinates}
\label{sec:S9}

\subsection{Maps that preserve the grading are orthogonal and linear}
\label{sec:S9-grading}

Let $E$ be a diffeomorphism of the standardized mapping space, acting on
densities at fixed bath variables by
$(\cU_Ef)(z)=j(z)\,f(D(z))$ with $D=E^{-1}$ and $j=|\det\partial D/\partial z|$,
so that normalization is preserved.  Suppose that $\cU_E$ maps $\cV_0$ onto
$\cV_0$ and $\cV_1$ onto $\cV_1$.  The first condition reads
$j(z)\,\varphi(D(z))=\kappa\,\varphi(z)$ with a constant $\kappa$; the second
requires $j(z)\,\varphi(D(z))\,D_a(z)$ to be $\varphi(z)$ times a polynomial of
first degree, for each component $a$.  Dividing the second by the first,
every $D_a(z)$ is a polynomial of first degree, so $D(z)=\mathsf Az+\bm b$ is
affine and $j=|\det\mathsf A|$ is constant.  The first condition then requires
$|\mathsf Az+\bm b|^2=|z|^2+{\rm const}$ for all $z$, whence
$\mathsf A^{\mathsf T}\mathsf A=\one$ and $\mathsf A^{\mathsf T}\bm b=0$,
so $\bm b=0$.  A map that preserves the two lowest Hermite spaces is therefore
an orthogonal linear map, and it preserves volume and every $\cV_K$.
Conversely, a nonlinear map cannot preserve both of the two lowest Hermite
spaces, and with them it loses the Hermite orders on which the closure of
Sec.~\Mref{sec:raising} rests.

\subsection{Unitary changes of the subsystem basis}

If the map must also preserve the Poisson bracket, so that the equations of
motion keep their form, $\mathsf A$ lies in
$O(2N)\cap Sp(2N,\mathbb R)$, which is the group $U(N)$ acting on the
amplitudes of Eq.~\eqref{eq:w} as $w\mapsto Uw$.  By Eq.~\eqref{eq:Pi-w},
$\sum_{\lambda\lambda'}h_{\lambda\lambda'}w_\lambda^*w_{\lambda'}$ and
$\sum_{\lambda\lambda'}\Pi_{\lambda\lambda'}\rho_W^{\lambda'\lambda}
=2\,w^\dagger\rho_W w-\Tr\rho_W$ keep their form with
$h\to UhU^\dagger$ and $\rho_W\to U\rho_WU^\dagger$: the map is a change of the
subsystem basis.  The operators $i\cL_m$ and $i\cL_m^{\PB}$ are both
covariant under it, and since the Gaussian inner product is invariant
under orthogonal maps, so is the norm of \Meq{eq:eps}, and with it the defect
rate $\mathcal D_4$.  No choice of
subsystem basis, and no linear change of mapping coordinates that keeps the
Hermite grading and the Poisson bracket, reduces the fourth-order defect.

This covariance was checked numerically with the production operator code for
\numword{\covNRot} changes of basis, \numword{\covNComplex} of them complex.  For
$\covNRand$ random Hermitian Hamiltonians the transformed rotation and force
matrices equal the conjugated originals to $\covOperator$; the induced map on
coefficients mixes no Hermite orders, to $\covOrderMix$, obeys the group
property to $\covGroup$ and maps the physical sector onto itself to
$\covFrame$.  A packet started at the avoided crossing was propagated for
$\covSteps$ steps to $t=\covT$ on a $\covGrid$ grid in the original and in
each transformed basis; the two calculations, which differ by
$\covSep$ in relative norm before the inverse transformation, agree after it
to $\covProp$.

\subsection{A nonlinear map that preserves the physical sector}

The grading, not the physical sector, is what the argument of Sec.~\ref{sec:S9-grading}
constrains.  The common-phase map $w\mapsto e^{i\alpha w^\dagger w}w$ is
nonlinear for $\alpha\ne0$, symplectic, and preserves $w^\dagger w$ and every
$w_\lambda w^*_{\lambda'}$; it therefore leaves $\varphi$, every
$\Pi_{\lambda\lambda'}$ and $H_m$ unchanged and maps the physical sector onto
itself.  It does not preserve $\cV_1$, so it does not retain the Hermite grading
used in the general closure argument.  Maps of this kind change
nothing physical; they show that the rigidity concerns the representation
used to close the dynamics, not the physical content.

\bibliography{references}

%% file: paper_values.tex
\newcommand{\TTPForty}{1500}
\newcommand{\TTPOneHundred}{600}
\newcommand{\TTPSeventy}{857.5}
\newcommand{\TTPTen}{6000}
\newcommand{\TTPThirty}{2000}
\newcommand{\TTPTwenty}{3000}
\newcommand{\clCtlEFour}{6.65}
\newcommand{\clCtlETwo}{7.49}
\newcommand{\clCtlT}{150}

\newcommand{\clFrame}{8.3\times10^{-11}}
\newcommand{\clGrid}{256\times128}
\newcommand{\clModes}{46}
\newcommand{\clModesFour}{35}
\newcommand{\clPZero}{40}
\newcommand{\clT}{1500}
\newcommand{\cohFocMax}{0.050}
\newcommand{\cohFocMin}{0.0022}

\newcommand{\cohGridMax}{1.3\times10^{-8}}
\newcommand{\cohPbmeOverSemMin}{6.8}

\newcommand{\cohPbmePOneHundred}{0.0062}

\newcommand{\cohPbmePTen}{0.060}

\newcommand{\cohPeakMax}{0.45}
\newcommand{\cohPeakMin}{0.30}

\newcommand{\cohQclePOneHundred}{1.7\times10^{-5}}

\newcommand{\cohQclePTen}{0.0074}

\newcommand{\cohRatioMax}{370}
\newcommand{\cohRatioMin}{7.3}
\newcommand{\cohSemMax}{0.0011}
\newcommand{\cohSemMin}{9.4\times10^{-4}}

\newcommand{\cohTPOneHundred}{0.21}
\newcommand{\cohTPSeventy}{0.23}

\newcommand{\cohTPTwenty}{1.5\times10^{-4}}

\newcommand{\dhDt}{1}
\newcommand{\dhDtPSeventy}{0.5}

\newcommand{\dhPeakPTen}{0.12}

\newcommand{\dhPeakRangeHi}{0.027}
\newcommand{\dhPeakRangeLo}{0.019}
\newcommand{\dhRangeMom}{1.33}
\newcommand{\dhSplitErr}{7\times10^{-5}}
\newcommand{\eGridFinMax}{7.4\times10^{-10}}
\newcommand{\eGridMaxAll}{4.6\times10^{-9}}

\newcommand{\eGridPTen}{2.7\times10^{-5}}

\newcommand{\ePbmePOneHundred}{0.0035}

\newcommand{\ePbmePTen}{0.10}

\newcommand{\ePbmeRangeMom}{15}

\newcommand{\ePbmeTabPForty}{3.03(7)\times10^{-2}}
\newcommand{\ePbmeTabPOneHundred}{3.5(10)\times10^{-3}}
\newcommand{\ePbmeTabPSeventy}{1.07(8)\times10^{-2}}
\newcommand{\ePbmeTabPTen}{1.047(3)\times10^{-1}}
\newcommand{\ePbmeTabPThirty}{3.22(8)\times10^{-2}}
\newcommand{\ePbmeTabPTwenty}{5.14(5)\times10^{-2}}

\newcommand{\eQclePOneHundred}{7.3\times10^{-6}}

\newcommand{\eQclePTwenty}{0.0054}
\newcommand{\eQcleTabPForty}{8.3\times10^{-4}}
\newcommand{\eQcleTabPOneHundred}{7.3\times10^{-6}}
\newcommand{\eQcleTabPSeventy}{9.7\times10^{-5}}
\newcommand{\eQcleTabPTen}{1.4\times10^{-3}}
\newcommand{\eQcleTabPThirty}{9.0\times10^{-4}}
\newcommand{\eQcleTabPTwenty}{5.4\times10^{-3}}

\newcommand{\eStrangPOneHundred}{1.3\times10^{-5}}

\newcommand{\eStrangPTen}{4.7\times10^{-8}}

\newcommand{\focCloserCount}{five}
\newcommand{\focDiffMax}{0.0347}
\newcommand{\focDiffMin}{0.0047}
\newcommand{\focNSamples}{200\,000}

\newcommand{\focPopTabPForty}{\phantom{-}0.3113}
\newcommand{\focPopTabPOneHundred}{\phantom{-}0.8531}
\newcommand{\focPopTabPSeventy}{\phantom{-}0.4354}
\newcommand{\focPopTabPTen}{-0.0941}
\newcommand{\focPopTabPThirty}{\phantom{-}0.6228}
\newcommand{\focPopTabPTwenty}{\phantom{-}0.0463}

\newcommand{\ghAdPopPTen}{0.0223}
\newcommand{\ghFrameMax}{3.3\times10^{-10}}

\newcommand{\ghGridPForty}{1024\times512}
\newcommand{\ghGridPOneHundred}{2048\times1024}
\newcommand{\ghGridPSeventy}{2048\times1024}
\newcommand{\ghGridPTen}{4096\times2048}
\newcommand{\ghGridPThirty}{1024\times512}
\newcommand{\ghGridPTwenty}{4096\times1024}
\newcommand{\ghHoursPForty}{0.45}
\newcommand{\ghHoursPOneHundred}{3.8}
\newcommand{\ghHoursPSeventy}{2.8}
\newcommand{\ghHoursPTen}{108}
\newcommand{\ghHoursPThirty}{0.5}
\newcommand{\ghHoursPTwenty}{13}
\newcommand{\ghMHigh}{12}
\newcommand{\ghMLow}{8}

\newcommand{\ghOverGridMax}{2.0}
\newcommand{\ghOverGridMin}{0.8}

\newcommand{\ghPboxPForty}{[18,62]}
\newcommand{\ghPboxPOneHundred}{[60,140]}
\newcommand{\ghPboxPSeventy}{[38,102]}
\newcommand{\ghPboxPTen}{[-30,50]}
\newcommand{\ghPboxPThirty}{[8,52]}
\newcommand{\ghPboxPTwenty}{[-20,60]}
\newcommand{\ghPopPForty}{0.299563}
\newcommand{\ghPopPOneHundred}{0.851828}
\newcommand{\ghPopPSeventy}{0.433134}
\newcommand{\ghPopPTen}{0.017007}
\newcommand{\ghPopPThirty}{0.641319}
\newcommand{\ghPopPTwenty}{0.062997}
\newcommand{\ghRboxPForty}{[-45,45]}
\newcommand{\ghRboxPOneHundred}{[-35,35]}
\newcommand{\ghRboxPSeventy}{[-35,35]}
\newcommand{\ghRboxPTen}{[-45,45]}
\newcommand{\ghRboxPThirty}{[-45,45]}
\newcommand{\ghRboxPTwenty}{[-45,45]}

\newcommand{\ghSubstepsPForty}{618}
\newcommand{\ghSubstepsPOneHundred}{1147}
\newcommand{\ghSubstepsPSeventy}{906}
\newcommand{\ghSubstepsPTen}{7761}
\newcommand{\ghSubstepsPThirty}{699}
\newcommand{\ghSubstepsPTwenty}{1925}
\newcommand{\ghTraceMax}{3.0\times10^{-9}}

\newcommand{\gridDtHalfMax}{1.5\times10^{-13}}

\newcommand{\gridDtLarge}{0.5}

\newcommand{\gridDtSmall}{0.25}

\newcommand{\gridSpatialMax}{1.6\times10^{-10}}

\newcommand{\krylovTol}{1\times10^{-10}}
\newcommand{\nModesGH}{11}
\newcommand{\nQuarticModes}{31}

\newcommand{\numUncHigh}{7.4\times10^{-10}}

\newcommand{\numUncPTen}{1.5\times10^{-5}}

\newcommand{\parCtlEOdd}{2.73}
\newcommand{\parCtlEOddFinal}{5.71}

\newcommand{\parModes}{35}
\newcommand{\parModesOdd}{24}
\newcommand{\pbAdPopPTen}{-0.0840}
\newcommand{\pbAdSemPTen}{0.0003}
\newcommand{\pbBlockSize}{5\,000}
\newcommand{\pbDtChange}{4.2\times10^{-6}}

\newcommand{\pbDtOverSem}{0.006}

\newcommand{\pbNSamples}{1\,200\,000}
\newcommand{\pbNSamplesMin}{1\,000\,000}

\newcommand{\pbPopPTen}{-0.0877}

\newcommand{\pbPopTabPForty}{\phantom{-}0.3298(7)}
\newcommand{\pbPopTabPOneHundred}{\phantom{-}0.8484(10)}
\newcommand{\pbPopTabPSeventy}{\phantom{-}0.4224(8)}
\newcommand{\pbPopTabPTen}{-0.0877(3)}
\newcommand{\pbPopTabPThirty}{\phantom{-}0.6091(8)}
\newcommand{\pbPopTabPTwenty}{\phantom{-}0.0116(5)}

\newcommand{\pbSemPTen}{0.0003}

\newcommand{\psCellPPThirty}{0.344}

\newcommand{\psCellRPThirty}{0.352}

\newcommand{\psGridPeakLowMax}{1\times10^{-6}}
\newcommand{\psGridPeakPOneHundred}{0.0036}
\newcommand{\psGridPeakPSeventy}{1.4\times10^{-4}}

\newcommand{\psGridSameMeasures}{3\times10^{-5}}

\newcommand{\psLtwoQcleWorseCohList}{20\text{ and }30}
\newcommand{\psLtwoQcleWorseTwoList}{20\text{ and }70}

\newcommand{\psMomMinQcleTwoPTwenty}{-0.038}
\newcommand{\psMomOverFloorMin}{8}

\newcommand{\psMomPeakQTwoPTwenty}{0.019}
\newcommand{\psMomQcleHighMax}{0.022}
\newcommand{\psMomQcleOneMax}{0.18}

\newcommand{\psMomQcleOnePThirty}{0.10}

\newcommand{\psMomQcleTwoPOneHundred}{0.0029}

\newcommand{\psMomQcleTwoPThirty}{0.070}
\newcommand{\psMomQcleTwoPTwenty}{3.0}
\newcommand{\psMomRatioMax}{15}
\newcommand{\psMomRatioMin}{1.2}
\newcommand{\psNegQMax}{0.002}

\newcommand{\psNegQcleOneMax}{0.098}
\newcommand{\psNegQcleOneMin}{0.043}

\newcommand{\psNegQcleTwoPTwenty}{0.19}

\newcommand{\psRkAfterMax}{2\times10^{-8}}
\newcommand{\psRkList}{30, 70\text{ and }100}

\newcommand{\psRkResidMax}{0.0031}
\newcommand{\psShiftHighList}{70\text{ and }100}
\newcommand{\psShiftPOneHundred}{0.20}
\newcommand{\psShiftPSeventy}{0.41}
\newcommand{\psShiftPThirty}{2.4}
\newcommand{\psShiftPTwenty}{5.9}

\newcommand{\psSplitPbmePOneHundred}{1.66}
\newcommand{\psSplitPbmePSeventy}{0.78}
\newcommand{\psSplitPbmePThirty}{4.04}

\newcommand{\psSplitQPOneHundred}{0.05}
\newcommand{\psSplitQPSeventy}{0.18}
\newcommand{\psSplitQPThirty}{3.19}
\newcommand{\psSplitQclePOneHundred}{0.05}
\newcommand{\psSplitQclePSeventy}{0.17}
\newcommand{\psSplitQclePThirty}{3.37}

\newcommand{\psWidthPbmeOnePThirty}{2.14}

\newcommand{\psWidthPbmeTwoPThirty}{1.58}

\newcommand{\qAdPopPTen}{0.0290}

\newcommand{\qDtProd}{0.125}

\newcommand{\qNRProdHigh}{16384}
\newcommand{\qNRProdLow}{8192}

\newcommand{\qPopPForty}{0.298728}
\newcommand{\qPopPOneHundred}{0.851821}
\newcommand{\qPopPSeventy}{0.433231}
\newcommand{\qPopPTen}{0.018436}
\newcommand{\qPopPThirty}{0.640418}
\newcommand{\qPopPTwenty}{0.057594}
\newcommand{\qUncMax}{5.1\times10^{-8}}

\newcommand{\qcleOverNumMin}{99}

\newcommand{\qcleOverQMin}{140}
\newcommand{\ratioMax}{480}
\newcommand{\ratioMin}{9.5}
\newcommand{\ratioPForty}{36}
\newcommand{\ratioPOneHundred}{470}
\newcommand{\ratioPSeventy}{110}
\newcommand{\ratioPTen}{73}
\newcommand{\ratioPThirty}{36}
\newcommand{\ratioPTwenty}{9.5}
\newcommand{\ratioTimeMin}{4.2}
\newcommand{\scanBound}{1.7\times10^{-5}}
\newcommand{\scanBoundC}{0.06}
\newcommand{\scanCList}{0.005, 0.0075, 0.015, 0.03, 0.06}
\newcommand{\scanGradSpan}{1.07}
\newcommand{\scanPeakRange}{1.07}

\newcommand{\scanRespMax}{0.030}

\newcommand{\sigRPForty}{0.25}
\newcommand{\sigRPOneHundred}{0.1}
\newcommand{\sigRPSeventy}{0.143}
\newcommand{\sigRPTen}{1}
\newcommand{\sigRPThirty}{0.333}
\newcommand{\sigRPTwenty}{0.5}

\newcommand{\strangHalveMax}{4.03}
\newcommand{\strangHalveMin}{3.99}

\newcommand{\strangOverQclePOneHundred}{1.8}
\newcommand{\strangScanErr}{6.2\times10^{-6}}